\documentclass[fleqn,usenatbib]{mnras}

\usepackage[T1]{fontenc}

\DeclareRobustCommand{\VAN}[3]{#2}
\let\VANthebibliography\thebibliography
\def\thebibliography{\DeclareRobustCommand{\VAN}[3]{##3}\VANthebibliography}

\usepackage{graphicx}	
\usepackage{amsmath}	
\usepackage{amssymb}	

\usepackage[usenames,dvipsnames]{xcolor} 

\usepackage{makecell}

\newcommand{\app}[1]{Appendix~\ref{sec:#1}}

\newcommand{\fig}[1]{Figure~\ref{fig:#1}}
\renewcommand{\sec}[1]{Section~\ref{sec:#1}}
\newcommand{\tab}[1]{Table~\ref{tab:#1}}
\newcommand{\App}[1]{Appendix~\ref{sec:#1}}

\newcommand{\eagle}{\mbox{\sc{Eagle}}}

\newcommand{\flares}{\mbox{\sc Flares}}

\newcommand{\deltasfs}{\mbox{$\Delta \mathrm{SFS}$}}

\title[FLARES: Early Passive Galaxies]{First Light And Reionisation Epoch Simulations (FLARES) VIII. The Emergence of Passive Galaxies at $z \geqslant 5$}

\author[C. C. Lovell et al.]{Christopher C. Lovell$^{1,2,3}$,\thanks{E-mail: christopher.lovell@port.ac.uk}
Will Roper$^{3}$,
Aswin P. Vijayan$^{4,5}$,
Louise Seeyave$^{3}$,\newauthor
Dimitrios Irodotou$^{6}$,
Stephen M. Wilkins$^{3}$,
Christopher J. Conselice$^{7}$,
Flaminia Fortuni$^{8}$,
\newauthor
Jussi K. Kuusisto$^{3}$,
Emiliano Merlin$^{8}$,
Paola Santini$^{8}$,
Peter Thomas$^{3}$
\\
$^{1}$Institute of Cosmology and Gravitation, University of Portsmouth, Burnaby Road, Portsmouth, PO1 3FX, UK\\
$^{2}$Centre for Astrophysics Research,  School of Physics, Engineering \& Computer Science, University of Hertfordshire, \\Hatfield AL10 9AB, UK\\
$^{3}$Astronomy Centre, University of Sussex, Falmer, Brighton BN1 9QH, UK\\
$^{4}$Cosmic Dawn Center (DAWN) \\
$^{5}$DTU-Space, Technical University of Denmark, Elektrovej 327, DK-2800 Kgs. Lyngby, Denmark \\
$^{6}$Department of Physics, University of Helsinki, Gustaf Hällströmin katu 2, FI-00014, Helsinki, Finland\\
$^{7}$Jodrell Bank Centre for Astrophysics, University of Manchester, Oxford Road, Manchester, UK\\
$^{8}$INAF Osservatorio Astronomico di Roma, Via Frascati 33, 00078 Monteporzio Catone, Rome, Italy\\
}

\date{Accepted XXX. Received YYY; in original form ZZZ}

\pubyear{2022}

\begin{document}
\label{firstpage}
\pagerange{\pageref{firstpage}--\pageref{lastpage}}
\maketitle

\begin{abstract}
Passive galaxies are ubiquitous in the local universe, and various physical channels have been proposed that lead to this passivity.
To date, robust passive galaxy candidates have been detected up to $z \leqslant 5$, but it is still unknown if they exist at higher redshifts, what their relative abundances are, and what causes them to stop forming stars.
We present predictions from the First Light And Reionisation Epoch Simulations (\flares), a series of zoom simulations of a range of overdensities using the \eagle\ code.
Passive galaxies occur naturally in the \eagle\ model at high redshift, and are in good agreement with number density estimates from HST and early JWST results at $3 \leqslant z \leqslant 5$.
Due to the unique \flares\ approach, we extend these predictions to higher redshifts, finding passive galaxy populations up to $z \sim 8$.
Feedback from supermassive black holes is the main driver of passivity, leading to reduced gas fractions and star forming gas reservoirs.
We find that passive galaxies at $z \geqslant 5$ are not identified in the typical UVJ selection space due to their still relatively young stellar populations, and present new rest--frame selection regions.
We also produce mock NIRCam and MIRI fluxes, and find that significant numbers of passive galaxies at $z \geqslant 5$ should be detectable in upcoming wide surveys with JWST.
Finally, we present JWST colour distributions, with new selection regions in the observer--frame for identifying these early passive populations.
\end{abstract}



\section{Introduction}

The tight relation between stellar mass and star formation rate (SFR), known as the star-forming sequence or `main sequence' \citep{daddi_multiwavelength_2007,noeske_star_2007} is well established, and present out to high redshift \citep{rodighiero_multiwavelength_2014,speagle_highly_2014,schreiber_herschel_2015}.
Whilst a significant fraction of galaxies lie on this relation at all epochs, there is significant scatter at fixed stellar mass \citep{matthee_origin_2017,corcho-caballero_single_2020,katsianis_specific_2021} as a result of varying processes which promote or inhibit star formation on a range of spatial and temporal scales \citep{tacchella_confinement_2016,iyer_diversity_2020}.
When a galaxy ceases to form stars it falls off the relation completely, and is described as being `passive' or `quiescent'.\footnote{We note that we do not distinguish between `passive', `quenched' or `quiescent' adjectives to describe galaxies in this article.}
At $z = 0$ a well established passive population exists, that correlates with both environmental density and stellar mass \citep{peng_mass_2010,vulcani_comparing_2010,paccagnella_slow_2016,darvish_cosmic_2017}, and these passive galaxies tend to have spheroidal or elliptical morphologies \citep{kauffmann_dependence_2003}.

A number of physical processes can contribute to the cessation of star formation in a galaxy \citep{man_star_2018}.
Active galactic nuclei (AGN) feedback, leading to the heating and expelling of gas from a galaxy, has been inferred observationally \citep{crenshaw_mass_2003}, and is necessary to match the high-mass end of the stellar mass function in models \citep{croton_many_2006,di_matteo_energy_2005,springel_black_2005}.
Feedback from evolved stars can also lead to similar effects \citep{ciotti_winds_1991,merlin_formation_2012}, and is invoked to match the number densities of lower mass galaxies below the knee of the halo mass function.
Galaxy assembly is a hierarchical process, and mergers, both major and minor, can have both a positive and negative impact on star formation \citep{van_der_wel_major_2009}, through compaction of the gas reservoir and transformation of the structure and morphology of the remnant \citep{naab_atlas3d_2014,lagos_quantifying_2018}.
Other mechanisms include starvation or strangulation of the gas supply  \citep{keres_how_2005,peng_strangulation_2015,feldmann_argo_2015}, a hot circum-galactic medium from virial shocking \citep{dekel_galaxy_2006}, and ram pressure stripping in a hot intra-cluster medium \citep{simpson_quenching_2018}.
Whilst all of these processes are expected to play a part at low-$z$, at higher redshift this is less clear.
Dense cluster environments do not exist at $z > 2$ \citep{overzier_realm_2016}, and so many environmentally dependent quenching processes, such as ram pressure stripping, cannot operate.
Despite this, passive galaxies have been detected at $z > 2$ \citep[\textit{e.g.}][]{merlin_chasing_2018,merlin_red_2019,carnall_timing_2020,shahidi_selection_2020,valentino_quiescent_2020}.
Passive number densities are negatively correlated with redshift, with the highest redshift confirmed candidates currently detected at $z \sim 5$ with a predicted number density of $\sim 10^{-5} \; \mathrm{Mpc^{-3}}$ \citep{merlin_red_2019,carnall_surprising_2023}.
It is unclear if these are some of the very first passive galaxies in the Universe, or whether populations of passive galaxies exist at even higher redshifts; indeed, early JWST results suggest this may be the case \citep{looser_discovery_2023}.
However, if their number densities continue to fall with increasing redshift then (coupled with their intrinsic faintness) it is unlikely that they would have been detected in HST observations to date.\footnote{\cite{merlin_red_2019} found a very high redshift passive candidate at $z = 6.7$ that passed their strict photometric selection, but could not confirm with spectra or FIR/sub-mm data.}

These early passive galaxies may have very different properties to those detected at lower redshift.
This may present an opportunity: the high redshift universe presents a unique laboratory to study quenching processes in essentially pristine galaxies with simple star formation histories, without many of the degeneracies present at lower redshift.
It is also expected that many passive objects detected at high redshift will start forming stars again in their future, due to the abundance of gas available for star formation in the early Universe \citep{rudnick_deep_2017,gobat_unexpectedly_2018,aravena_alma_2019,whitaker_high_2021,magdis_interstellar_2021}.
The passive `phase' observed may be only a few tens of Myr long before star formation restarts; the timescale of passivity is therefore an important consideration, particularly when comparing samples of passive galaxies at different epochs.
Different observational tracers are also sensitive to star formation over different timescales -- from a few tens of Myr for H$\alpha$ and UV emission lines, to hundreds of Myr in the NIR--FIR \citep{katsianis_evolution_2017,katsianis_high-redshift_2020}.

A passive galaxy exhibits a number of observational signatures compared to those that have recently or are actively forming stars.
One of the most common diagnostics is position in the rest--frame UVJ colour--colour plane.
UV colour distinguishes unobscured star formation from redder populations, but the addition of VJ colour allows the degeneracy between red, dust obscured starbursts and evolved (passive) populations to be broken \citep{labbe_irac_2005,wuyts_what_2007,williams_detection_2009}.
However, there are further degeneracies with metallicity \citep{worthey_comprehensive_1994} and dust \citep{cimatti_k20_2002,dunlop_systematic_2007} that place additional uncertainty on the colour selection of passive populations \citep{merlin_chasing_2018}.
The effect of these degeneracies at high redshift is unclear; metallicity estimates are typically made using strong line calibrations derived from lower redshift objects \citep{maiolino_re_2019}, and the dust reserviors in high-$z$ galaxies have been shown to be surprisingly large even at $z = 5$ \citep[\textit{e.g.}][]{fudamoto_normal_2021}.

The recent launch of JWST is already enabling the identification of passive galaxy candidates out to higher redshifts and at lower masses than was possible previously \citep{carnall_surprising_2023,marchesini_early_2023,perez-gonzalez_ceers_2023}.
However, at high redshift ($z > 3$) the rest--frame NIR is inaccessible to NIRCam, requiring templates derived from stellar population synthesis (SPS) models to infer the UVJ colour.
The MIRI instrument can probe these wavelengths at high redshift, but at much reduced sensitivity, limiting its use for passive galaxy identification.
It is therefore crucial to understand any uncertainties or biases in the SPS models that may lead to errors in the interpretation of high redshift passive populations.

On the theoretical side, a number of numerical models have been used to study the star forming sequence and its evolution, as well as the emergence of the passive galaxy population.
A general trend in almost all models, both semi-analytic and those employing full hydrodynamics, is that the normalisation of the star forming sequence is underpredicted at cosmic noon ($z \sim 2$) compared to observational estimates \citep{lu_semi-analytic_2014,katsianis_relation_2016}.
There is still debate on the causes of this discrepancy; it may be due to selection and measurement biases in observationally inferred values using different tracers, particularly due to uncertainties in dust attenuation effects \citep[\textit{e.g.}][]{katsianis_evolution_2017}.
It may also be due to biases in the way that properties are measured in simulations; \cite{katsianis_high-redshift_2020} demonstrated that measuring properties using a more sophisticated forward modelling pipeline alleviates much of the tension at $1 < z < 4$.
In \eagle\ \citep{schaye_eagle_2015}, \cite{furlong_evolution_2015} showed a similar negative offset in the normalisation of the star forming sequence at $z = 2$, but good agreement with passive fractions.
Similarly, \cite{donnari_star_2019} showed that in Illustris-TNG \citep{pillepich_simulating_2018} the passive fractions at $z \leqslant 2$ are in good agreement, but they also demonstrated the sensitivity to star formation averaging timescales and aperture.
In \cite{donnari_quenched_2021,donnari_quenched_2021-1} they extended this analysis out to $z = 3$, and explored the contributions to the passive population in both centrals and satellites from internal feedback as well as environmental processes.
\cite{merlin_red_2019} presented predictions from a number of theoretical models up to $z \leqslant 5$, including both \eagle\ and \textsc{Illustris-TNG}, as well as the original \textsc{Illustris} and the \textsc{Simba} simulation \citep{dave_simba:_2019}.
They found significant variation in the passive number densities, up to 1.5 dex at $z = 3$, but found good agreement with observational constraints at $z > 3$ in \eagle, though a slight underestimate at lower redshifts, whereas other models show the opposite redshift behaviour.
Simba in particular agrees with the data better at lower redshifts, and shows very good agreement with the observed UVJ diagram at $z \leqslant 2$ \citep{akins_quenching_2022}.

In this paper we explore predictions for the passive galaxy population in the Epoch of Reionisation (EoR; $z \geqslant 5$) using the First Light And Reionisation Epoch Simulations \citep[\flares,][]{lovell_first_2021,vijayan_first_2021}, a series of `zoom' hydrodynamic simulations using the \eagle\ code \citep{schaye_eagle_2015,crain_eagle_2015}.
Due to its unique approach to region selection, \flares\ allows us to study very rare objects that would not be produced in similar resolution periodic simulations, due to their limited volume.
This makes it an ideal simulation for studying rare, passive populations in the early universe, with sufficient resolution to explore the details of the physics leading to their passivity.
We have also shown that the normalisation and shape of the star forming sequence in \flares\ is in good agreement with observational estimates at $z \geqslant 5$ \citep{lovell_first_2021,dsilva_unveiling_2023}, giving us confidence in the model for self-regulated star formation in this redshift regime.

We also explore predictions for the colour evolution of passive galaxies, both in the observer- and rest-frame.
By combining \flares\ with simple stellar population models we can evaluate the detectability of passive galaxies during this early stage of galaxy evolution, using typical rest--frame UVJ selection criteria as well as both current and upcoming JWST observations using NIRCam and MIRI photometry.

The paper is laid out as follows.
In \sec{methods} we describe the \eagle\ and \flares\ simulations, and our definitions and fiducial choices for aperture, SFR and quiescence.
\sec{number_densities} presents our predictions for volume normalised and surface number densities.
In \sec{causes} we explore the causes of quiescence by looking at individual galaxies as well as population demographics.
And in \sec{obs} we describe our forward modelling approach, and show how rest--frame and observer--frame colours can be used to select passive objects at high-$z$.
Finally, we discuss our results and present our conclusions in \sec{conclusions}.
We assume a Planck year 1 cosmology \citep[$\Omega_{\mathrm{m}} = 0.307$, $\Omega_{\Lambda} = 0.693$, $h = 0.6777$;][]{planck_collaboration_planck_2014}, consistent with that used in \flares\ and \eagle, and a \cite{chabrier_galactic_2003} Initial Mass Function (IMF) throughout.

\section{Simulations and Definitions}
\label{sec:methods}

\subsection{The \textsc{EAGLE} Model and Simulations}

The \eagle\ simulation project is a series of hydrodynamic cosmological simulations, with varying resolutions and box sizes \citep{schaye_eagle_2015,crain_eagle_2015}.
\eagle\ uses a smoothed particle hydrodynamics (SPH) code based on a heavily modified version of \textsc{P-Gadget-3}, last described in \cite{springel_simulations_2005}.
The hydrodynamic suite, collectively known as `\textsc{Anarchy}' \citep[described in detail in][]{schaye_eagle_2015,schaller_eagle_2015}, consists of the \cite{hopkins_general_2013} pressure-entropy SPH formalism, an artifical viscosity switch \citep{cullen_inviscid_2010}, an artificial conductivity switch \citep[e.g.][]{price_modelling_2008}, the \cite{wendland_piecewise_1995} $C^{2}$ smoothing kernel with 58 neighbours, and the \cite{durier_implementation_2012} time-step limiter.
Subgrid model prescriptions for radiative cooling and photo-heating \citep{wiersma_effect_2009}, star formation \citep{schaye_relation_2008}, stellar evolution and mass loss \citep{wiersma_chemical_2009}, feedback from star formation \citep{dalla_vecchia_simulating_2012}, black hole growth and AGN feedback \citep{springel_black_2005,booth_cosmological_2009,rosas-guevara_impact_2015} are then included.
The free parameters of these models were calibrated to the $z=0.1$ galaxy stellar mass function, the gas mass--halo mass relation, the stellar mass -- black hole mass relation, and disc galaxy sizes.
The model is in good agreement with a number of observables at low-redshift not considered in the calibration \citep[e.g.][]{furlong_evolution_2015,trayford_colours_2015,lagos_molecular_2015}.

In this study we make use of the fiducial Reference periodic simulation volume, which is 100 Mpc on a side, run from $z = 127$ to $0$.
We utilise the lower redshift snapshots ($z \leqslant 5$) to cover the redshift range not simulated by \flares, which we discuss in the next section.

\subsection{The First Light And Reionisation Epoch Simulations}
\label{sec:flares}

The First Light And Reionisation Epoch Simulations \citep[\flares,][]{lovell_first_2021,vijayan_first_2021} are a series of zoom simulations using the \eagle\ physics model.
Full details are provided in \cite{lovell_first_2021,vijayan_first_2021}; here we briefly review the relevant aspects for this study.
\flares\ consists of 40 spherical `zoom' regions, each $14 h^{-1} \mathrm{Mpc}$ in radius.
Each region is selected at $z = 4.67$ from a very large $(3.2 \, \mathrm{cGpc})^3$ parent dark matter only (DMO) simulation \citep{barnes_redshift_2017}.
These regions are then run from $z = 127$ to $z = 4.67$, capturing the epoch of reionisation.
Regions are selected in order to sample an unprecedentedly large range of environments, particularly overdense regions where the most massive, extreme galaxies are expected to reside.
In order to build composite distribution functions the regions are combined using their overdensity as a weighting, allowing the combined zooms to approximate a much larger volume than that explicitly simulated.
These distribution functions cover a much larger dynamic range than that achievable with typical periodic simulations at the same resolution.

\flares\ uses an identical resolution to the fiducial 100 cMpc \eagle\ Reference simulation introduced above, with a dark matter particle mass of $m_{\mathrm{dm}} = 9.7 \times 10^6\, \mathrm{M}_{\odot}$, initial gas particle mass of $m_{\mathrm{g}} = 1.8 \times 10^6\, \mathrm{M}_{\odot}$, and gravitational softening length of $2.66\, \mathrm{ckpc}$ at $z\geq2.8$.
\flares\ employs the AGNdT9 parameter configuration of \eagle, which includes a higher value for $C_{\mathrm{visc}}$ (controlling the sensitivity of the BH accretion rate to the angular momentum of the gas) and a higher gas temperature increase from AGN feedback, $\Delta T$ (which leads to fewer, more energetic feedback events).
Together these parameter changes lead to a closer match with observational constraints on the hot gas properties in groups and clusters \citep{barnes_cluster-eagle_2017}

The \eagle\ model has been shown to reproduce distribution functions at high-redshift reasonably well \citep{furlong_evolution_2015}, and this was extended to higher redshifts in \flares.
\cite{vijayan_first_2021,vijayan_first_2022} also demonstrated that \flares\ is able to reproduce the observed UV luminosity function at $z \geqslant 5$, and galaxy sizes are in good agreement with HST constraints at $z \sim 5$ \citep{roper_first_2022}.
We have also explored galaxy populations at the redshift `frontier' \citep[$z > 10$;][]{wilkins_first_2022}, the evolution of galaxy colours with redshift \citep{wilkins_first_2022}, and the behaviour of star formation and metal enrichment histories during the epoch of reionisation \citep{wilkins_first_2023-1}.

In \sec{obs} we model the observational signatures of passive galaxies; full details on the forward modelling pipeline applied to our simulations are provided there.

\begin{figure}
    \centering
    \includegraphics[width=20pc]{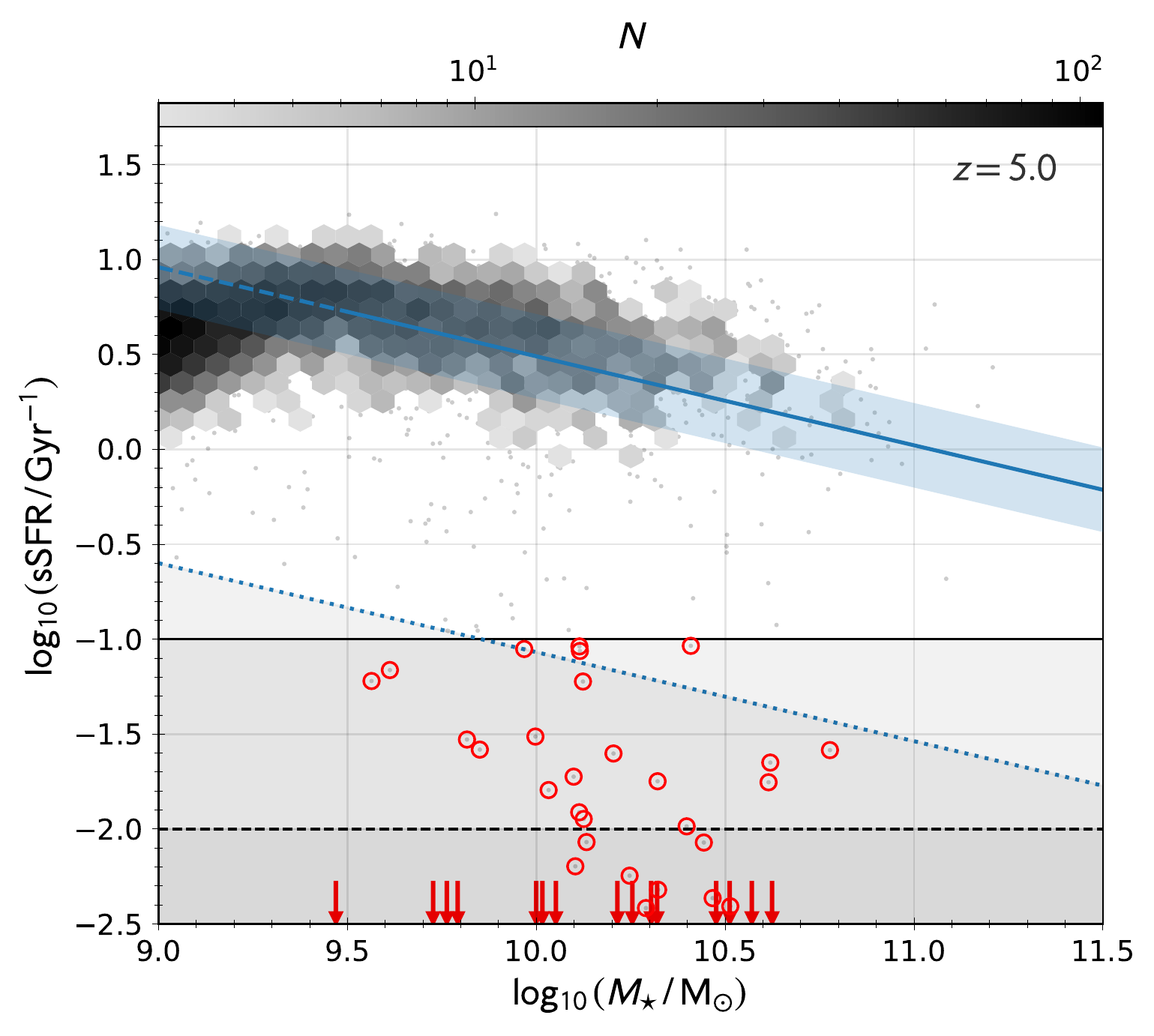}
    \caption{
      The star forming sequence in \flares\ at $z = 5$.
      All galaxies are shown as grey points, binned where there are more than 3 points.
      Individual galaxies that satisfy any of the passive selections are highlighted with red circles, and those with sSFR's below the plot limits are shown with red arrows
      The solid and dashed black lines show the $\mathrm{log_{10}(sSFR \,/\, Gyr^{-1})} > [-1, -2]$ selection thresholds, respectively.
      The solid blue line shows the fit to the star-forming sequence at $M_{\star} \,/\, \mathrm{M_{\odot}} > 10^{9.5}$, and the dotted blue line shows the $-1.5 \; \mathrm{dex}$ offset selection.
      The extrapolation of the fit to lower masses is shown as a dashed blue line.
    }
    \label{fig:selection.sSFR}
\end{figure}

\subsection{Aperture and Timescale Definitions}
\label{sec:aperture_timescale}

In order to best match simulated properties with those inferred observationally, an important consideration is the aperture within which the property is defined.
\cite{donnari_star_2019} showed that this can have a large impact on the normalisation of the star forming sequence at $z = 0 - 2$, and in \cite{wilkins_first_2022} we showed that it also impacts low mass ($< 10^{9.7} \, M_{\star} \,/\, \mathrm{M_{\odot}}$) galaxy properties at $z = 5$.
Observations of passive galaxies at early times reveal very compact morphologies, $< 2.5 \, \mathrm{kpc}$ \citep{cimatti_gmass_2008,williams_progenitors_2014,merlin_red_2019}, and in \flares\ we found similarly compact intrinsic sizes across the whole galaxy population \citep{roper_first_2022}.
In \flares\ we measure both star formation rates and stellar masses in a range of aperture sizes, $R = [1, 3, 5, 10, 20, 30, 40, 50, 70, 100] \mathrm{pkpc}$.
For consistency with observational measurements, we measure SFRs, stellar masses and other particle based properties within the aperture closest to $4 \times R_{1/2,\star}$ \citep{merlin_red_2019,santini_emergence_2021}.

An additional consideration is the timescale over which the SFR is measured.
Observationally this is sensitive to the probe used \citep{kennicutt_star_2012,wilkins_recalibrating_2019}.
In \flares\ we can measure SFRs in two different ways.
The instantaneous SFR is the sum of the SFRs from each star forming gas particle, and encodes the state of the star forming gas in the galaxy at the point of observation.
Longer timescales can be approximated by summing the mass in stars formed over some time window.
At $z = 2$ \cite{donnari_star_2019,donnari_quenched_2021} found that the normalisation of the star forming sequence and the quenched fraction are highly sensitive to the choice of timescale, particularly at low masses, whereas in \flares\ at $z = 5$ we found that the star forming sequence is relatively insensitive \citep{wilkins_first_2022}.
However, passive galaxies, particularly at low mass, may be more sensitive to this choice.
This is due to the limited resolution of the simulations - a single newly formed star particle has a mass of $10^6 \, \mathrm{M_{\odot}}$, larger than typical HII regions, which in a low-mass galaxy can move it directly from the passive to the star forming selection (when defined using specific star formation rate, see \sec{selection}).
We choose a fiducial SFR averaging timescale of 50 Myr (to capture both recent and intermediate age episodes of star formation), and explore the dependence on timescale where relevant.

\subsection{Definitions of Passivity}
\label{sec:selection}

Typically in observational data, the colour in rest--frame UVJ space is used to define a passive region, excluding galaxies reddened by dust \citep[\textit{e.g.}][]{williams_detection_2009}.
In \sec{obs} we explore whether our galaxies cover these regions defined at lower redshifts, and the completeness and purity of the galaxy population within these selection regions.
However, in the simulation we can directly measure the SFR and stellar mass, so the position on the star forming sequence can be determined explicitly.
In this study we use four different cuts in specific SFR (sSFR; the SFR of a galaxy normalised by its stellar mass) to define quiescence.
We use two fixed cuts, defined as
\begin{align}
  \mathrm{log_{10}(sSFR \,/\, Gyr^{-1})} < [-1,-2] \;\;.
\end{align}
A galaxy with stellar mass $M_{\star} \,/\, \mathrm{M_{\odot}} = 10^{10}$ is then defined as quiescent when it has a $\mathrm{SFR} < [1,0.1] \; \mathrm{M_{\odot} \, yr^{-1}}$, respectively.

We also employ two cuts that evolve with redshift.
The first measures the distance of a galaxy in the sSFR--$M_{\star}$ plane from the ridge of the star-forming sequence.
This is well defined at high redshift, and can be fit linearly in the high-mass regime.
In this definition, all galaxies $\leqslant 1.5 \; \mathrm{dex}$ below the star-forming sequence are selected as passive.
This selection is less sensitive to the evolution of the sSFR distribution \citep{wilkins_first_2023-1}.
We will refer to this selection as $\Delta \mathrm{SFS}$.
Finally, the fourth cut employs a time dependent cut in sSFR \citep{gallazzi_charting_2014, pacifici_evolution_2016, carnall_surprising_2023},
\begin{align}
  \mathrm{log_{10}(sSFR \,/\, Gyr^{-1})} < 0.2 \,/\, t_{\mathrm{obs}}(z) \;\;.
\end{align}
where $t_{\mathrm{obs}}$ is the age of the universe at redshift $z$.
We will refer to this as the \textit{evolving selection}.
We will show results assuming all four cuts throughout the article, to enable comparison with observational results assuming different thresholds.

\fig{selection.sSFR} shows three of the four selections at $z = 5$ in \flares.
The more liberal sSFR cut and the distance from the star forming sequence produce similar numbers of passive galaxies, though the latter has a lower threshold at high stellar masses.
The evolving selection gives a $\mathrm{log_{10}\; sSFR \,/\, Gyr^{-1}}$ threshold of $\sim -0.5$ at $z = 8$, $\sim -0.75$ at $z = 5$, and falls to $\sim -1.5$ by $z = 1$.
Above a stellar mass of $10^{9} \, \mathrm{M_{\odot}}$, the fixed sSFR cuts produce 43 and 23 passive galaxies in all the \flares\ regions combined, the $\Delta\mathrm{SFS}$ cut produces 41, and the evolving cut 226.

\section{Number Densities}
\label{sec:number_densities}

We begin by presenting the \flares\ predictions for the number density of quiescent galaxies as a function of redshift.
As described in \sec{flares}, \flares\ is a series of zoom simulations, not a single periodic simulation.
These regions were selected by their overdensity from a large parent box; to obtain composite number densities we must weight each region by its relative abundance, as defined by that overdensity, in the parent simulation \citep[see][for details]{lovell_first_2021}.
The outcome of this procedure is a weight, $w$, for each region, which can be used to obtain the composite quiescent fraction,
\begin{align}
  N_{Q} \,/\, \mathrm{Mpc^{-3}} = \sum_{i} \, w_{i} \, n_{i,q} \;\;,
\end{align}
where $i$ is the region index, $n_{i,q} \,/\, \mathrm{Mpc^{-3}}$ is the volume density of quiescent galaxies in that region, and $w_{i}$ is the weight, where $\sum_i w_{i} = 1$.
A similar approach can be used to obtain surface number densities.

We also compare to lower redshift results from the periodic \eagle\ simulations.
In particular, we measure the quiescent number density in the flagship fiducial $(100 \; \mathrm{Mpc})^3$ Reference simulation.
Whilst this uses different parameters for the AGN feedback compared to the AGNdT9 variant of the model (used in \flares), the large box size leads to a better probe of the high mass end of the galaxy stellar mass function compared to the $(50 \; \mathrm{Mpc})^3$ AGNdT9 periodic simulation \citep[see][for details]{schaye_eagle_2015,crain_eagle_2015},  particularly at high-$z$.

\subsection{Volume Normalised Number Densities}

\begin{figure}
    \centering
    \includegraphics[width=20pc]{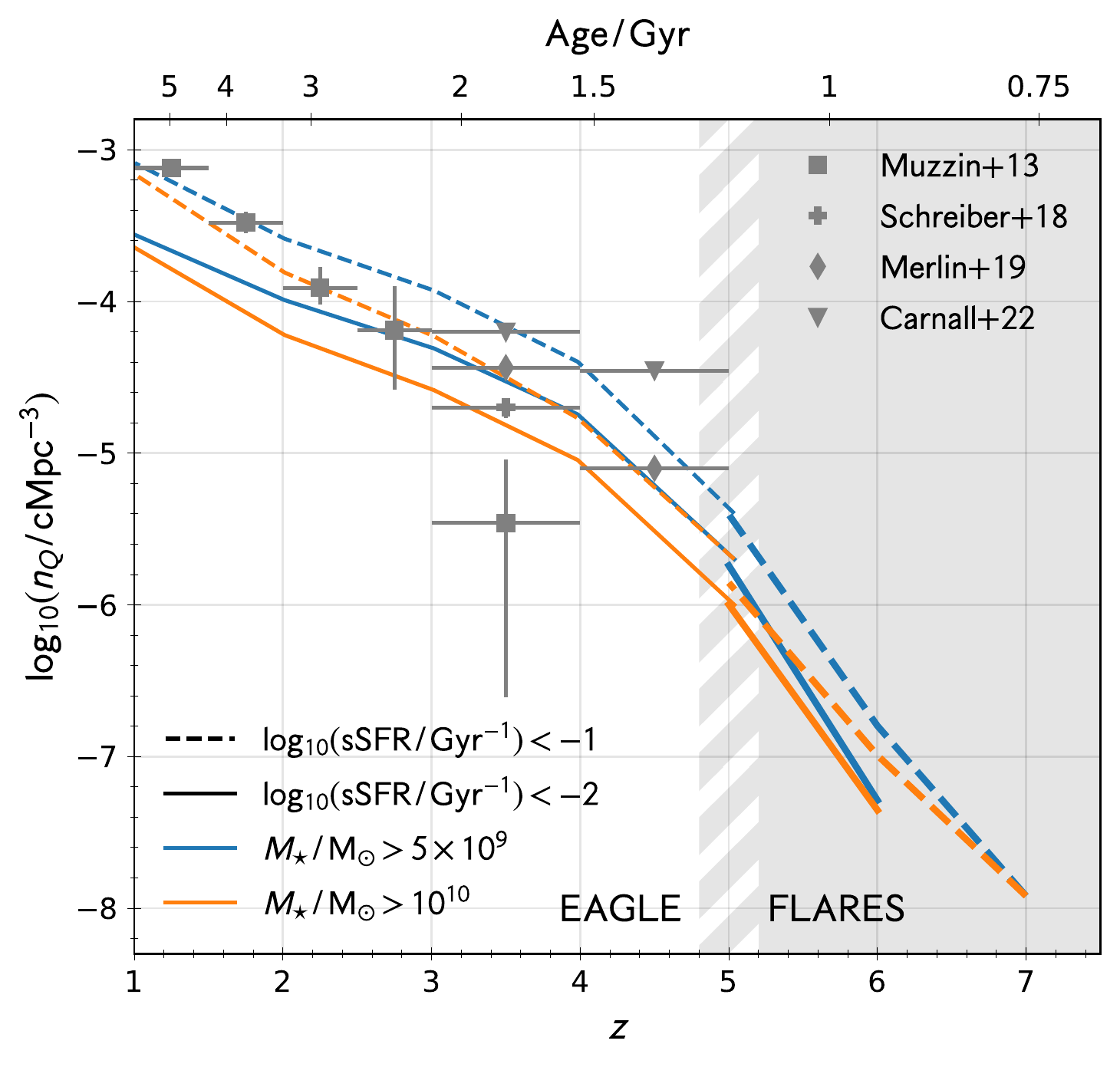}
    \includegraphics[width=20pc]{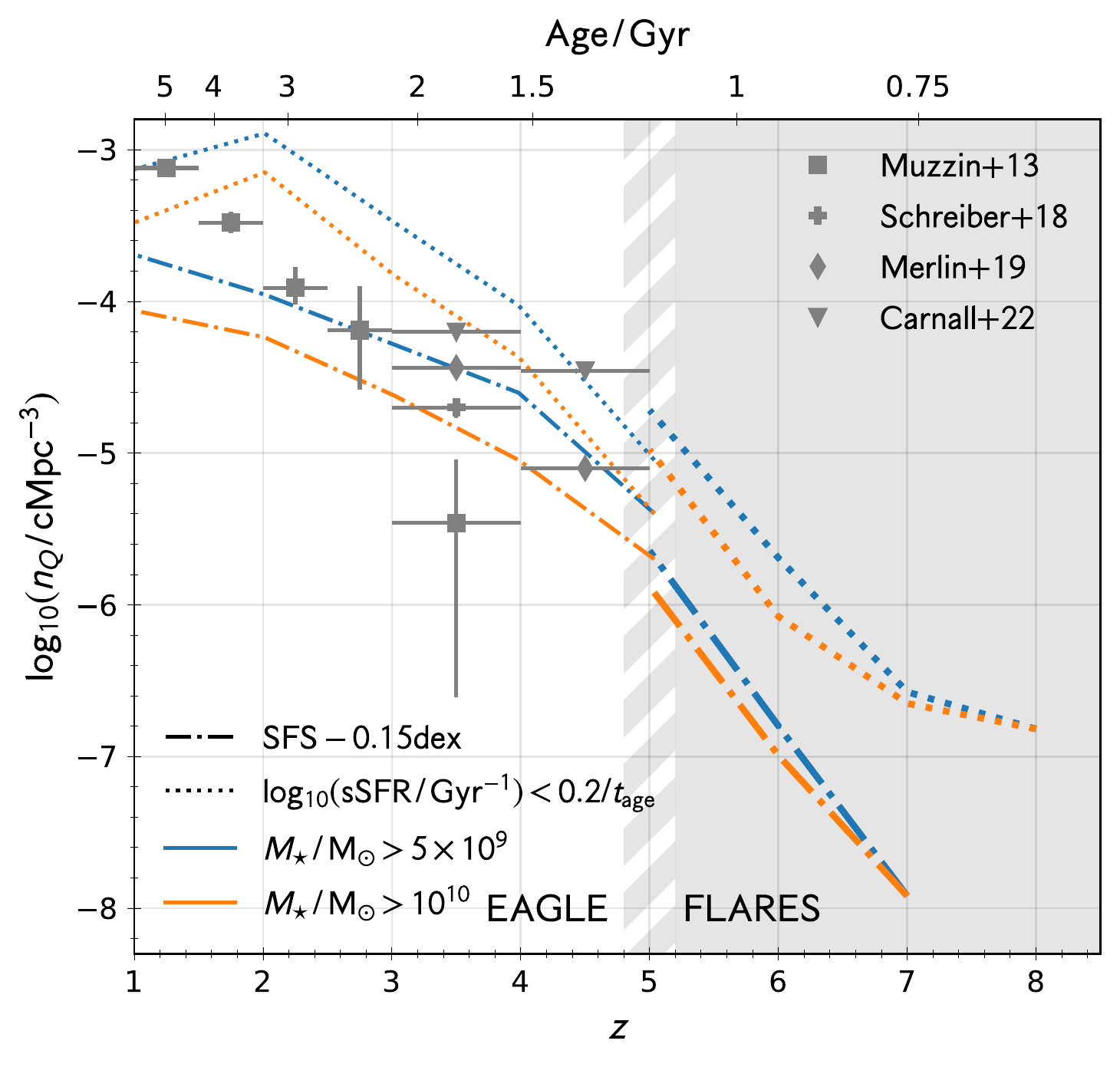}
    \caption{
        Volume normalised quiescent galaxy number density in \flares\ ($z \geqslant 5$) and the fiducial periodic \eagle\ simulation (Ref; $z \leqslant 5$).
        Two stellar mass selections, $M_{\star} \,/\, \mathrm{M_{\odot}} > [5 \times 10^{9}, \; 10^{10}]$, are shown with blue and orange lines, respectively.
        We also show a number of recent observational constraints at $z \leqslant 5$ from \protect\cite{muzzin_evolution_2013,schreiber_near_2018,merlin_red_2019,carnall_surprising_2023}.
        \textit{Top panel:} two passive selections, $\mathrm{log_{10}(sSFR \,/\, Gyr)} > [-1,-2]$ (dashed and solid lines, respectively).
        \textit{Bottom panel:} the $\Delta \mathrm{SFS}$ and evolving selections are shown by the dotted and dashed-dotted lines, respectively.
    }
    \label{fig:q_fraction}
\end{figure}

\fig{q_fraction} shows the quiescent number density in \flares\ at $z \geqslant 5$, as well as the predicted number densities from \eagle\ at $1 \leqslant z \leqslant 5$.
These values are also provided in \tab{number_densities}.
The quantitative number densities are highly sensitive to the sSFR and stellar mass limit chosen.
To demonstrate we show four sSFR cuts (detailed in \sec{selection}), and two stellar mass limits, $M_{\star} \,/\, \mathrm{M_{\odot}} > [5 \times 10^{9}, \; 10^{10}]$.
For all selections there is a negative correlation with redshift, except for \deltasfs, which turns over at the lowest redshifts.
At $z < 4$ this is approximately proportional to $\propto (1+z)^{-0.43}$, whereas at $z > 4$ this steepens, to $\propto (1+z)^{-1.1}$.
The number densities for \textsc{Flares} and the periodic \textsc{Eagle} volume are in reasonably good agreement where they overlap at $z \sim 5$, which verifies that our weighting procedure is correct, as well as allowing us to compare the quiescent number densities over an unprecedented range in redshift.
Any remaining offsets can be attributed to parameter differences between the \eagle\ and \flares\ models.

\begin{table}
    \centering
    \caption{Volume normalised number densities, $\mathrm{log_{10}}\, n_{q} \,/\, \mathrm{Mpc^3}$, from \flares\ and \eagle\ for a range of redshifts ($1 \leqslant z \leqslant 7$, shown in \fig{q_fraction}). Two passive selections in terms of sSFR and two selections in stellar mass are provided.}
    \label{tab:number_densities}
    \begin{tabular}{ l|cccc }
    \hline
         & \multicolumn{4}{c}{\flares} \\
        $M_{\star} \,/\, \mathrm{M_{\odot}} >$ & \multicolumn{2}{c}{$5 \times 10^{9}$} & \multicolumn{2}{c}{$10^{10}$} \\
        $\mathrm{log_{10}\, \frac{sSFR}{Gyr^{-1}}} <$ & -1 & -2 & -1 & -2 \\
        z & & & & \\
      \hline
        7 & -7.92 & - & -7.92 & - \\
        6 & -6.80 & -7.29 & -6.99 & -7.35 \\
        5 & -5.40 & -5.74 & -5.85 & -6.00 \\
      \hline
    \end{tabular}
    \begin{tabular}{ l|cccc }
    \hline
         & \multicolumn{4}{c}{\eagle} \\
        $M_{\star} \,/\, \mathrm{M_{\odot}} >$ & \multicolumn{2}{c}{$5 \times 10^{9}$} & \multicolumn{2}{c}{$10^{10}$} \\
        $\mathrm{log_{10}\, \frac{sSFR}{Gyr^{-1}}} <$ & -1 & -2 & -1 & -2 \\
        z & & & & \\
      \hline
        5 & -5.40 & -5.70 & -5.70 & -6.00 \\
        4 & -4.40 & -4.74 & -4.77 & -5.05 \\
        3 & -3.93 & -4.31 & -4.23 & -4.59 \\
        2 & -3.59 & -3.99 & -3.81 & -4.22 \\
        1 & -3.09 & -3.56 & -3.16 & -3.65 \\
      \hline
    \end{tabular}
\end{table}

\fig{q_fraction} also shows a number of observational constraints at $z \leqslant 5$ \citep{muzzin_evolution_2013,schreiber_near_2018,merlin_red_2019,carnall_surprising_2023}.
These provide a test of the \textsc{Eagle} model in this intermediate-redshift regime.
As shown above, the quiescent number density is sensitive to the inferred stellar mass limit and the definition of quiescence, and as such care must be taken when comparing to the simulation results.
We highlight below which selections in \fig{q_fraction} are relevant for each study.
Quiescent galaxies are identified in most studies using a colour-colour selection, since this is not dependent on the SED fitting model (as long as the observations cover the photometric range in the rest--frame).
We convert all assumed IMFs to \cite{chabrier_galactic_2003}; whilst this does not affect the sSFR, it does change the lower stellar mass limit, which can affect the inferred number density.

\cite{merlin_red_2019} use a selection method, first developed in \cite{merlin_chasing_2018}, to select passive galaxies in the CANDELS fields.
They do SED fitting on $z>3$ sources with a library of models, assuming top--hat parametric star formation histories and a \cite{salpeter_luminosity_1955} IMF; when converted to a Chabrier IMF this leads to lower stellar masses by a factor of 0.63 \citep{madau_cosmic_2014}, giving a lower stellar mass limit of $7.94 \times 10^{9} \; \mathrm{M_{\odot}}$; this lies between the limits shown for \eagle\ and \flares\ in \fig{q_fraction}.
They consider as passive candidates those where the best-fit model is defined as passive (\textit{i.e.} $\mathrm{SFR}=0$ for a top-hat, which can best be compared to our $\mathrm{log_{10}(sSFR \,/\, Gyr^{-1})} < -2$ selection) with a probability >30\%, and where no star-forming solutions have a probability >5\%.
Their results are in good agreement with our lower sSFR predictions (solid lines) at $4 \leqslant z \leqslant 5$, and slightly overpredict at $3 \leqslant z \leqslant 5$, though they do not quote observational uncertainties, which would reduce any tension.

\cite{schreiber_near_2018} use a combined UVJ and template fitting selection, and a quiescent definition in the latter of $\mathrm{log_{10}\, sSFR < -0.82 \,/\, Gyr^{-1}}$, slightly higher than our most liberal fixed definition.
They select all galaxies with stellar masses $\geqslant 2 \times 10^{10} \mathrm{M_{\odot}}$, assuming a \cite{chabrier_galactic_2003} IMF, so their results are best compared to our liberal sSFR cut and higher mass limit (dashed orange line).
At $3 \leqslant z \leqslant 4$, the \eagle\ results are in good agreement with the inferred number densities.

\cite{muzzin_evolution_2013} use a UVJ selection criteria, and infer a \citep[IMF corrected]{kroupa_variation_2001} stellar mass limit of $10^{9.4} \, \mathrm{M_{\odot}}$.
They do not infer star formation rates for their sources, but the UVJ selection has been shown to select sources with sSFRs up to $\mathrm{log_{10}}\, \mathrm{sSFR < -0.6 \, Gyr^{-1}}$ \citep{straatman_substantial_2014}.
It is therefore appropriate to compare their predictions to our more liberal sSFR cut (dashed blue line).
At $z \sim 1$ their results are in good agreement with \eagle.
At higher redshifts the simulations \textit{over}predict the abundance of similar passive objects, however at these redshifts the \cite{muzzin_evolution_2013} results are highly incomplete.

Recently, \cite{carnall_surprising_2023} identified passive galaxy candidates in the first data from JWST NIRCam imaging in two redshift bins from $3 \leqslant z \leqslant 5$.
They use the redshift evolving sSFR criterion for passivity.
They also measure stellar masses using the Bagpipes SED fitting code, and find candidates with masses down to $10^{9.8} \, \mathrm{M_{\odot}}$, between our two mass selections.
We can compare these results to the dotted lines in \fig{q_fraction}.
Comparing to the lower stellar mass cut, the simulation results at $3 \leqslant z \leqslant 4$ are somewhat higher than the observations.
In the higher redshift bin, $4 \leqslant z \leqslant 5$, the observed number densities are in good agreement with those predicted by \eagle\ for the higher mass cut.
Recent results also suggest that many of the passive candidates in \cite{carnall_surprising_2023} contain active galactic nuclei \citep{kocevski_ceers_2023}; whilst many of these still have UVJ colours in the passive selection after removal of the AGN contribution, it's unclear how these AGN hosts affect the predicted number densities.

\begin{figure}
  \centering
  \includegraphics[width=20pc]{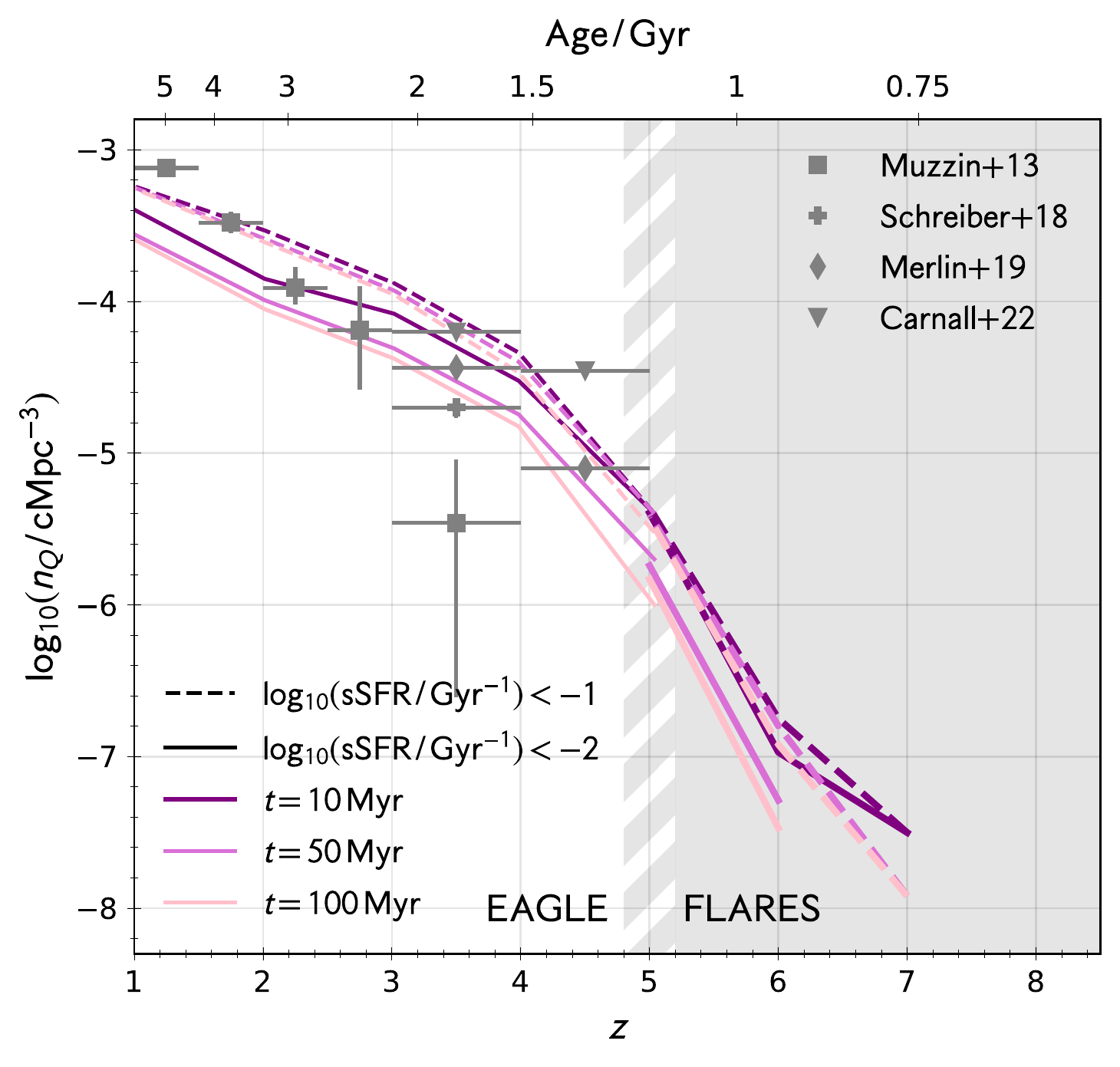}
  \caption{
      As for \fig{q_fraction}, but showing 10 Myr, 50 Myr and 100 Myr star formation rate timescale definitions (purple, orchid and pink lines, respectively).
      Showing all galaxies above $M_{\star} \,/\, \mathrm{M_{\odot}} > [5 \times 10^{9}, \; 10^{10}]$.
      Two fixed passive selections are shown, $\mathrm{log_{10}(sSFR \,/\, Gyr)} > [-1,-2]$ (dashed and solid lines, respectively).
  }
  \label{fig:q_fraction_timescale}
\end{figure}

Not shown in \fig{q_fraction} are the results from \cite{straatman_substantial_2014} and \cite{girelli_massive_2019}, who also both use a UVJ selection criteria.
They infer a stellar mass limit of $10^{10.6} \, \mathrm{M_{\odot}}$ (assuming a \citet{chabrier_galactic_2003} IMF), much higher than the studies described above.
When comparing directly to the number densities in \eagle\ given this mass limit we find a significant offset, with \eagle\ predicting far lower numbers of such massive, passive galaxies.
Part of this discrepancy can be explained by the drop off in the GSMF in \eagle\ for high stellar masses in the redshift range $1 < z < 5$ \citep{furlong_evolution_2015,santini_emergence_2021}.
This may partly be due to finite volume effects \citep[see][]{reed_halo_2007}.

In summary, observational studies are in significant tension with one another at these redshifts, however when looking at the details of the passive and stellar mass selection we find that results from \eagle\ at $z \leqslant 5$ are generally in good agreement.
Both the sSFR cut and lower stellar mass cut have a big impact on the predicted number densities, and we highlight the importance of comparing to simulations using identical cuts.

A final variable that can affect the quantitative number densities measured is the \textit{timescale} over which the star formation is measured, as discussed in \sec{aperture_timescale}.
Bursty star formation can lead to galaxies oscillating in and out of the passive selection if the timescale is short.
Such `mini-quenching' may be caused by different processes than those responsible for longer timescale quenching in the early Universe \citep{trussler_both_2020, donnari_quenched_2021-1,donnari_quenched_2021}.
\fig{q_fraction_timescale} shows the number densities of quiescent galaxies when measured over the most recent 10, 50 and 100 Myr of their star formation history.
It is clear that the number densities are higher for the shorter timescales, due to the capture of more of these mini-quenched systems.
There is also a dependence on the selection threshold employed.
However, over much of cosmic time the quantitative impact is relatively small ($< 0.05$ and $< 0.3$ dex for $\mathrm{log_{10}(sSFR \,/\, Gyr)} > [-1,-2]$, respectively), and peaks at $z \sim 5$ (0.6 dex for $\mathrm{log_{10}(sSFR \,/\, Gyr)} > -2$).

There are no confirmed constraints at $z > 5$ to directly compare with the \flares\ results; these are predictions that may be testable in upcoming wide field surveys, explored later in this section.
The existence of passive galaxies at $z > 5$, whilst rare, is a very interesting result.
If confirmed, it shows that feedback, potentially from supermassive black holes, begins to influence galaxy evolution just 1 Gyr after the big bang.

\subsection{Surface Number Densities}

\begin{figure}
  \centering
  \includegraphics[width=20pc]{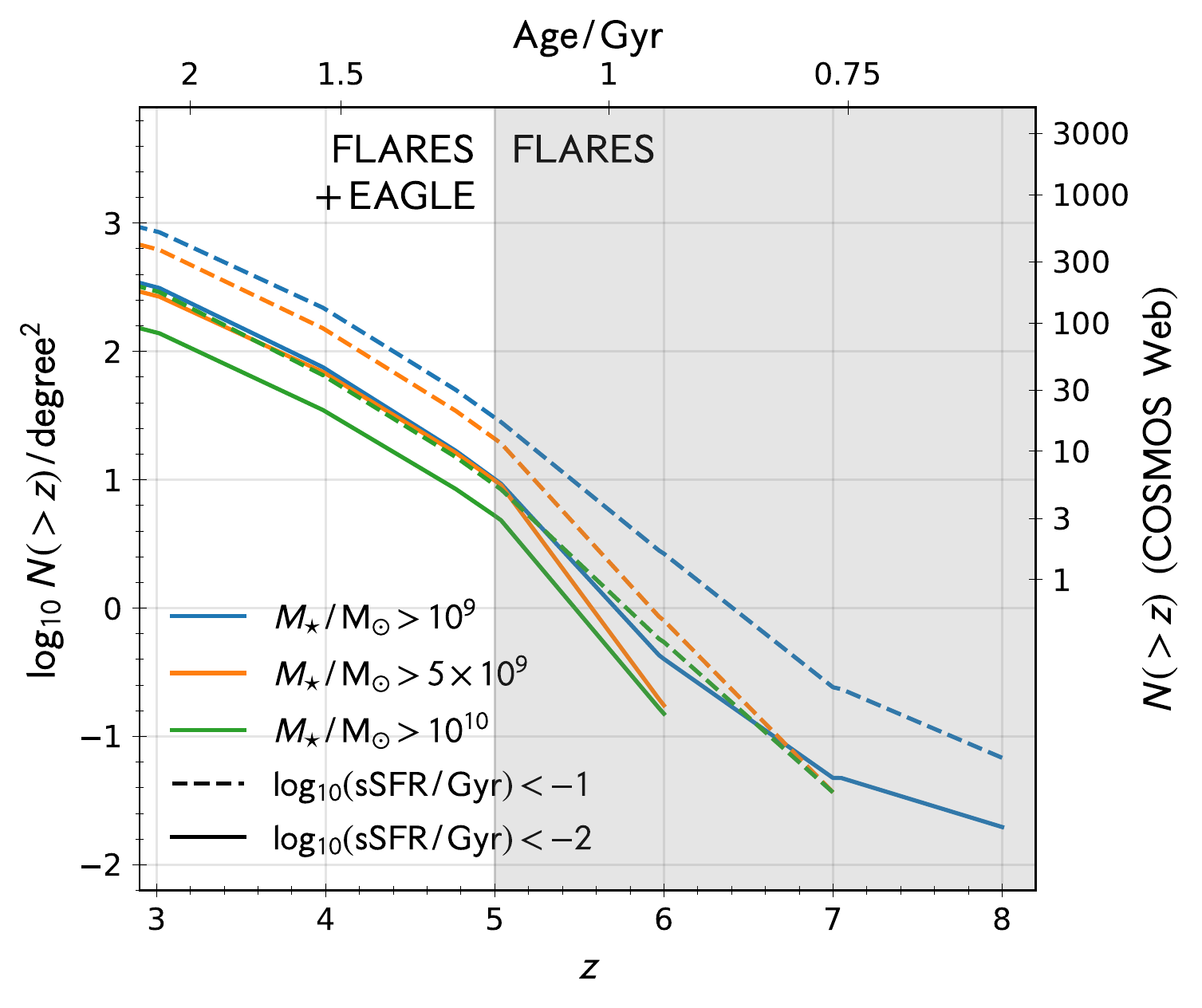}
  \includegraphics[width=20pc]{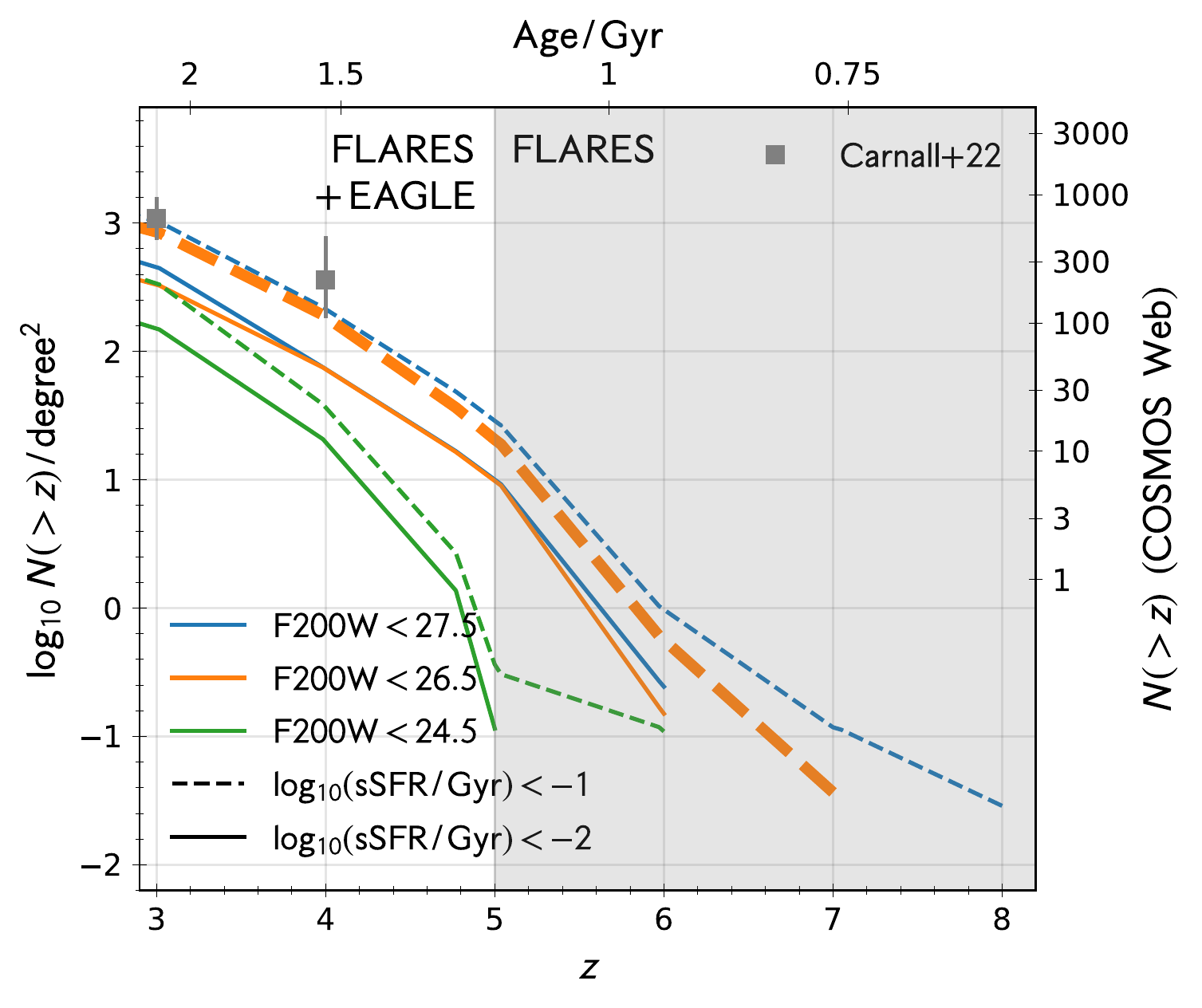}
  \caption{Cumulative surface densities ($> z$) of quiescent sources in \eagle\ and \flares\ combined.
  We show two passive definitions, $\mathrm{log_{10}(sSFR \,/\, Gyr^{-1})} > [-1,-2]$, with dashed and solid lines, respectively.
  \textit{Top:} three stellar mass selections are shown, $M_{\star} \,/\, \mathrm{M_{\odot}} > [10^{9}, \; 5 \times 10^{9}, \; 10^{10}]$ (blue, orange and green lines, respectively)
  \textit{Bottom:} three flux cuts are shown, $\mathrm{F200W} < [27.5, 26.5, 24.5]$ (blue, orange and green lines, respectively).
  The $\mathrm{F200W} < 26.5$ is highlighted in bold, and should be directly compared to the \protect\cite{carnall_surprising_2023} results, shown as the grey square points.
  }
  \label{fig:surface_densities}
\end{figure}

\begin{table}
    \centering
    \caption{Cumulative surface densities, $\mathrm{log_{10}} \; N(> z) \,/\, \mathrm{degree^2}$, from \flares\ and \eagle\ combined, for a range of redshifts ($3 \leqslant z \leqslant 8$, shown in \fig{surface_densities}). Two passive selections in terms of sSFR and three selections in F200W flux are provided.}
    \label{tab:surface_densities}
    \begin{tabular}{ l|cccccc }
    \hline
         & \multicolumn{6}{c}{\flares\ + \eagle} \\
        $\mathrm{F200W} <$ & \multicolumn{2}{c}{$27.5$} & \multicolumn{2}{c}{$26.5$} & \multicolumn{2}{c}{$24.5$}\\
        $\mathrm{log_{10}\, \frac{sSFR}{Gyr^{-1}}} <$ & -1 & -2 & -1 & -2 & -1 & -2 \\
        z & & & & & & \\
      \hline
        8 & -1.54 & - & - & - & - & - \\
        7 & -0.93 & - & -1.44 & - & - & - \\
        6 &  -0.01 &  -0.61 & -0.25 & -0.82 & -0.96 & - \\
        5 &  1.93 &  1.48 &  1.41 &  0.99 &  0.47 & 0.32 \\
        4 &  2.83 &  2.36 &  2.37 &  1.95 &  1.99 & 1.63 \\
        3 &  3.56 &  3.10 &  3.03 &  2.66 &  2.79 & 2.39 \\
      \hline
    \end{tabular}
\end{table}

In \fig{surface_densities} we show the predicted cumulative ($> z$) number of objects per square degree, for a number of physical and observational selection criteria.
These values are also provided in \tab{surface_densities}.
The trends with redshift for physical selection criteria (top panel) are similar to those shown in \fig{q_fraction}, however we additionally show predictions for a lower stellar mass limit of $M_{\star} \,/\, \mathrm{M_{\odot}} \geqslant 10^{9}$.
\flares\ predicts passive galaxies with such low stellar masses out to $z \sim 8$, considering both fixed sSFR selection criteria.

In the bottom panel of \fig{surface_densities} we again show the \cite{carnall_surprising_2023} results, this time comparing directly to their F200W < 26.5 selection.
We reiterate that their redshift-evolving selection is equivalent to $\mathrm{log_{10}(sSFR \,/\, Gyr^{-1})} < -1$ in the lower redshift bin ($3 \leqslant z \leqslant 4$), and more liberal in the higher redshift bin ($4 \leqslant z \leqslant 5$).
There is remarkable agreement with the \eagle\ results.

We also show predictions, on the right-hand axis, for the number of objects in the upcoming Cosmos Web survey \citep{kartaltepe_cosmos-webb_2021}.
This survey will cover 0.6 degree$^2$ with NIRCam photometry, and 0.2 degree$^2$ with MIRI.
We predict that, for a depth of F200W < 27.5, there will be approximately [18, 51] passive objects at $z \geqslant 5$ in the NIRCam survey area, assuming our two different fixed sSFR cuts.
These numbers drop off quite dramatically for more conservative flux cuts.

\begin{figure*}
    \centering
    \includegraphics[width=\textwidth]{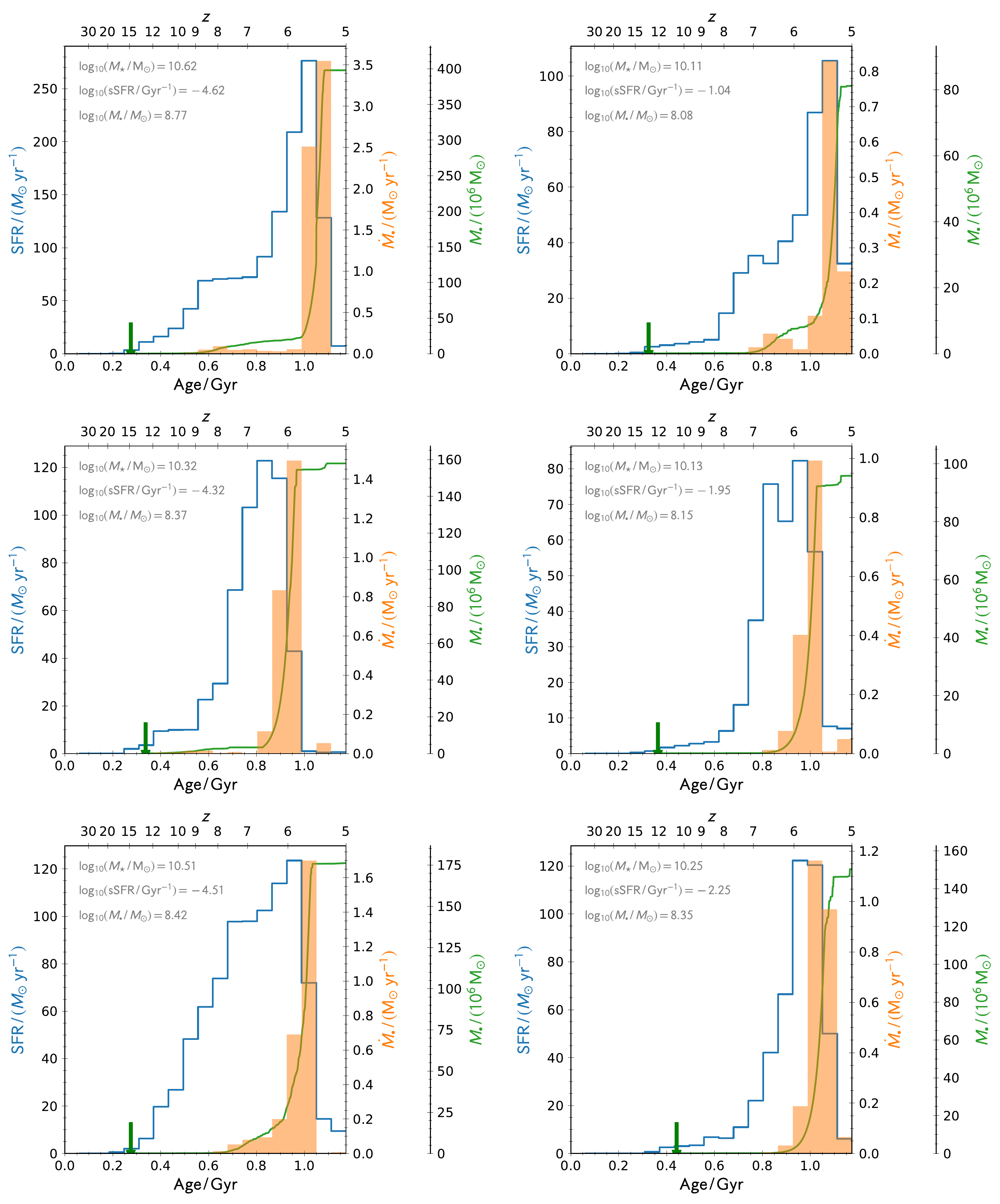}
    \caption{Individual passive galaxies from \flares\ at $z = 5$. 
    Each panel shows the SFH (blue, for the entire subhalo), black hole accretion history (orange), and central black hole mass (green). 
    Green arrows show the redshift at which the central black hole was seeded (see \sec{black_hole_mass} for details). 
    Each panel also details the stellar mass, sSFR and black hole mass of each galaxy at $z = 5$ measured within a 30 kpc aperture, as well as the \flares\ region they reside in and their ID.
    }
    \label{fig:SFH}
\end{figure*}

\begin{figure*}
    \centering
    \includegraphics[width=\textwidth]{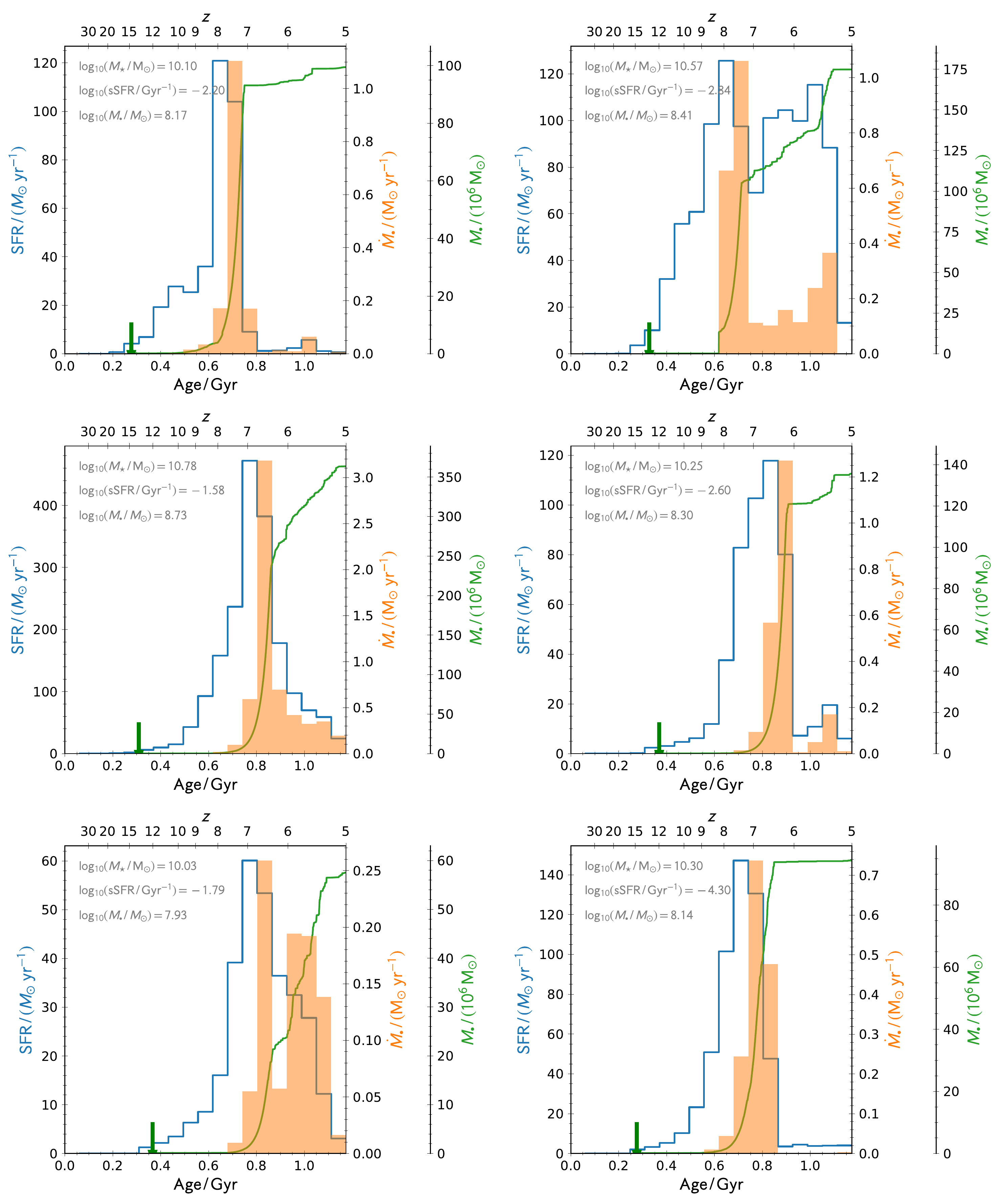}
    \caption{As for \fig{SFH}, showing 6 additional galaxies.}
    \label{fig:SFH_B}
\end{figure*}

\section{What causes passivity?}
\label{sec:causes}

So far we have looked at the volume normalised and surface number densities of passive objects in \eagle\ and \flares, and found good agreement in the $z \leqslant 5$ regime where observational constraints exist.
We have also made predictions for the number densities of passive populations at higher redshifts.
We now look in detail at individual passive galaxies selected at $z = 5$ from \flares, as well as the passive population as a whole at $z = 5$, to try to understand what causes the cessation of star formation in these high-$z$ objects.

\subsection{Individual Passive Galaxies}

Figures \ref{fig:SFH} \& \ref{fig:SFH_B} show the star formation history (SFH) and black hole accretion history for twelve passive galaxies selected from \flares\ at $z = 5$.\footnote{The SFH is shown for the entire subhalo, not just assuming the inner 30 kpc aperture, since star particles may move in and out of this aperture over time.}
These galaxies are all massive, $M_{\star} \,/\, \mathrm{M_{\odot}} > 10^{10}$, with $\mathrm{log_{10}(sSFR \,/\, Gyr^{-1})} < -1$\footnote{Measured within the 30 kpc  aperture at $z = 5$.}, and were seeded with black holes in the early universe ($z > 11$).
We note that black hole accretion here is measured using the change in black hole mass over a given interval.
It also includes the contribution from mergers, however we have investigated these examples in detail and found that all black hole mergers are between a massive central and black holes close to the seed mass; they therefore contribute negligibly to the black hole mass assembly at $z < 10$.

What is immediately clear in all these examples is the strong anti-correlation between black hole accretion and star formation rate.
In almost all examples, a single strong episode of black hole accretion injects significant energy into the ISM of the galaxy, and leads to the cessation of star formation.
The black hole accretion also then ceases, presumably due to a lack of non-turbulent gas available in the vicinity of the black hole after the feedback event.
Before this feedback event most galaxies have smoothly rising star formation histories, similar to those found in the wider galaxy population at $z = 5$ \citep{wilkins_first_2022}.
After this feedback event most galaxies have truncated SFHs, and the passive episode can last for up to 400 Myr or longer.
We have not performed a detailed exploration of rejuvenation of passive galaxies in this work, but we do see evidence for passive galaxies observed at $z = 6$ and $7$ that are star forming by $z=5$ in \flares.

\subsection{Passive Galaxy Demographics}
\label{sec:demographics}

\subsubsection{Black Hole Mass}
\label{sec:black_hole_mass}

\begin{figure}
    \centering
    \includegraphics[width=\columnwidth]{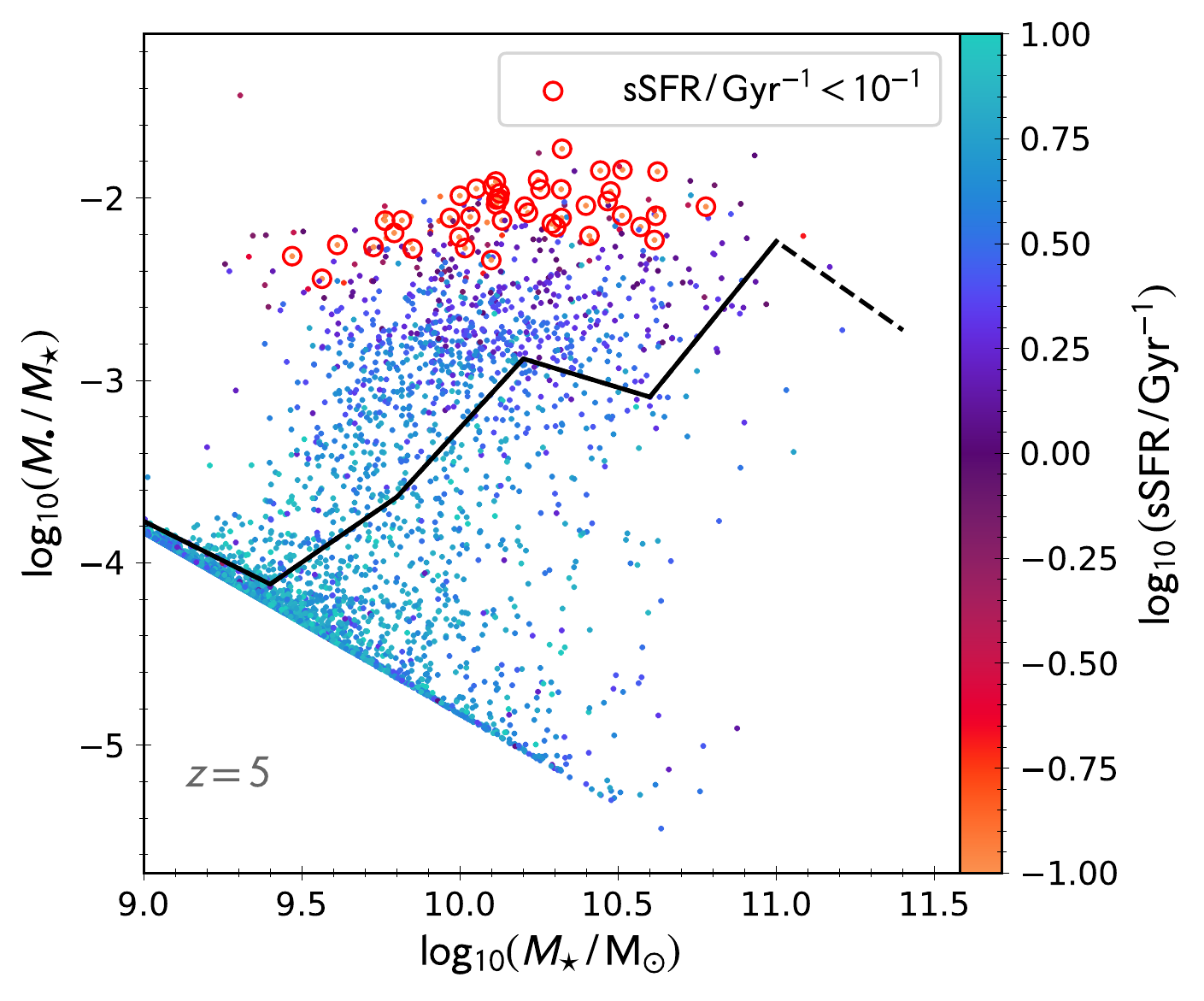}
    \caption{
      Specific black hole mass ($M_{\bullet} \,/\, M_{\star}$) against stellar mass at $z = 5$ for all \flares\ galaxies, coloured by sSFR.
      The weighted median is shown by the black line.
      Galaxies in our passive selection ($\mathrm{sSFR < 10^{-1} \,/\, Gyr^{-1}}$) are highlighted with red circles.
      }
    \label{fig:bh_mass}
\end{figure}

In the \eagle\ model, central black holes (BH) are seeded in all friends-of-friends (FOF) halos of mass $10^{10} \; h^{-1} \, \mathrm{M_{\odot}}$.
They then begin to accrete gas at a rate dependent on the mass of the BH, the density and temperature of the local gas, and the BHs relative velocity to that gas \citep{rosas-guevara_impact_2015}.
The BH then grows proportional to the rate of accretion, and injects thermal feedback in to the surrounding ISM.
This feedback energy is injected stochastically, and the magnitude is proportional to the accretion rate.
The mass of a black hole can then be thought of as a proxy for the total amount of feedback energy injected into a galaxy's ISM during its lifetime (ignoring BH mergers, which contribute negligibly at these redshifts).
Full details on the physics of black holes in the EAGLE model are provided in \cite{schaye_eagle_2015}.

\fig{bh_mass} shows the specific (central) black hole mass, $M_{\bullet} \,/\, M_{\star}$, against stellar mass, for both centrals and satellites.
Many galaxies have black holes close to the seed mass ($10^5 \, h^{-1} \, \mathrm{M_{\odot}}$) up to $M_{\star} \,/\, \mathrm{M_{\odot}} \sim 10^{10.5}$, which manifests as the diagonal line in the bottom left of the plot.
The majority of galaxies BHs grow rapidly in mass, beginning to set up the stellar-black hole mass relation seen at lower redshifts \citep{bower_dark_2017}.
Passive galaxies tend to lie above the locus of this relation; black holes in passive galaxies are more massive than the general population, at fixed stellar mass.
This suggests that black hole mass is an important driver of passivity, and agrees with the qualitative trends seen in Figures \ref{fig:SFH} \& \ref{fig:SFH_B}.
Interestingly, this is the case for all passive galaxies, regardless of whether they are centrals or satellites.

\subsubsection{Total and Star-Forming Gas Fractions}

\begin{figure}
    \centering
    \includegraphics[width=\columnwidth]{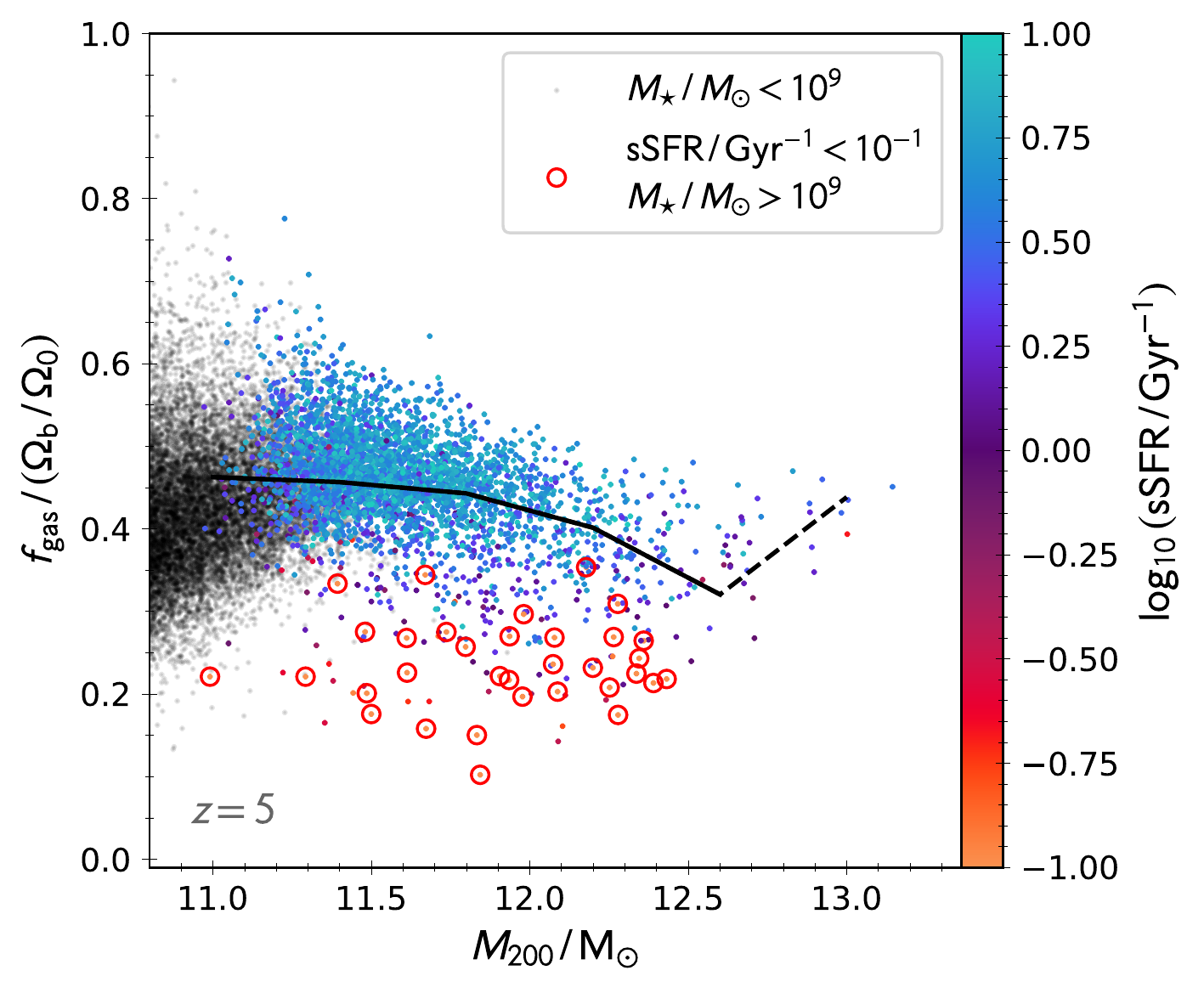}
    \includegraphics[width=\columnwidth]{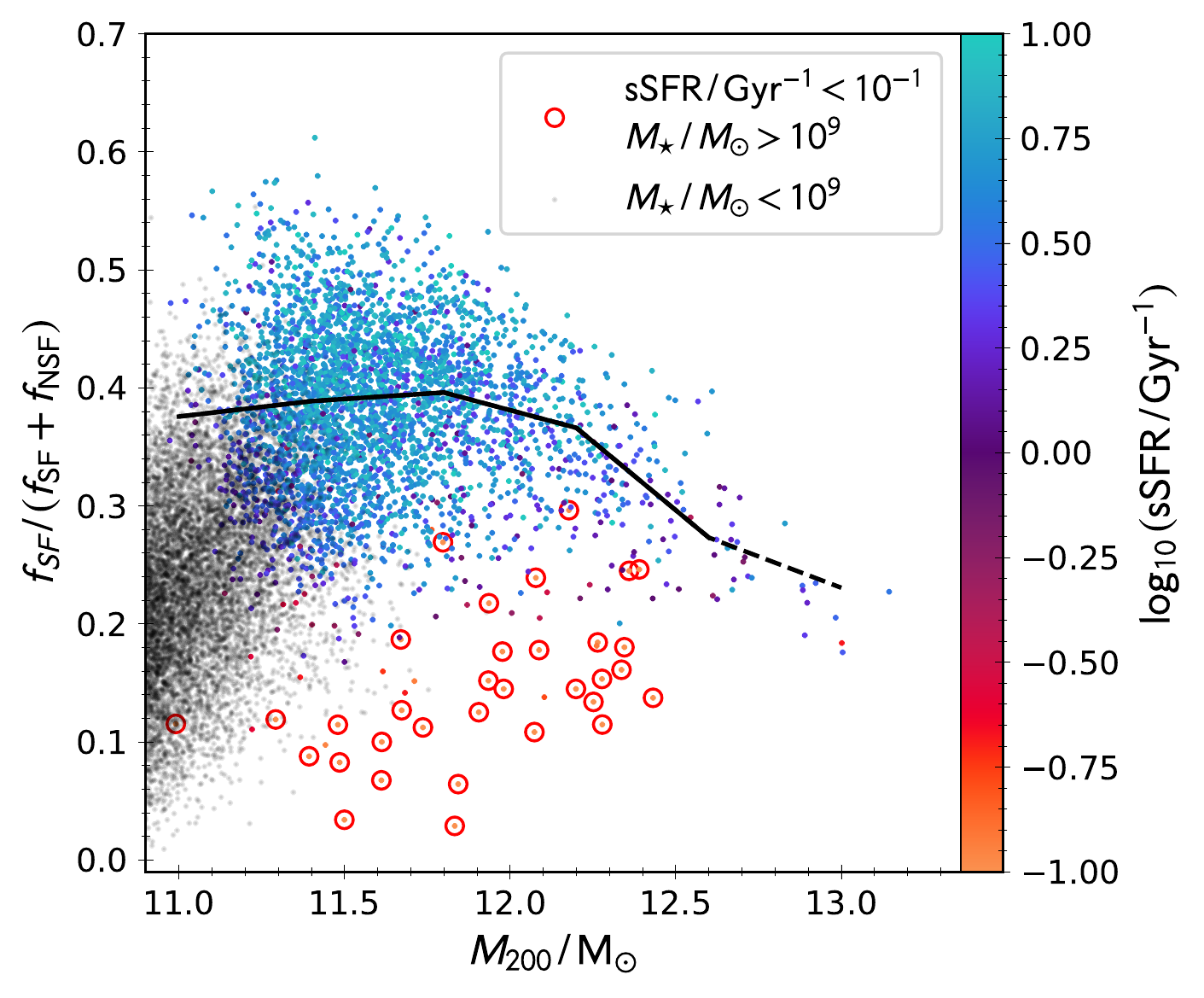}
    \caption{
        All \flares\ galaxies with $M_{\star} \,/\, \mathrm{M_{\odot}} > 9$ at $z = 5$ coloured by sSFR.
        The weighted median is shown by the black line, dashed where there are fewer than ten galaxies in a bin.
        Galaxies with $M_{\star} \,/\, \mathrm{M_{\odot}} < 9$ are coloured grey.
        Galaxies in our passive selection ($\mathrm{sSFR < 10^{-1} \,/\, Gyr^{-1}}$) are highlighted with red circles.
        \textit{Top:} Total gas fraction of central galaxies, normalised by the cosmic baryon fraction, against halo mass ($M_{200}$).
        \textit{Bottom:} Total star forming gas fraction against halo mass.
    }
    \label{fig:gas_fraction}
\end{figure}

The top panel of \fig{gas_fraction} shows the \textit{total} gas fraction of haloes (normalised by the cosmic baryon fraction) against halo mass in \flares\ at $z = 5$.
Contrary to the previous plots, we show this for the entire halo, and do not limit to observationally motivated apertures.
At high halo masses there is a reasonably tight relation centred on a gas fraction of $\sim 0.4$, but at lower masses the scatter increases considerably, with gas fractions higher than 0.8 in a handful of $M_{200} \,/\, \mathrm{M_{\odot}} = 10^{11}$ haloes.

Our passive selection has considerably lower total gas fractions than the overall population at fixed halo mass.
This suggests that the effect of black hole feedback in these passive galaxies, evidenced in the previous sections, is to reduce the overall gas fraction of the halo.
Since we are looking at the \textit{overall} halo, we can also infer that this gas has been ejected completely outside of the halo, and not just from the central regions where star formation is concentrated.
This highlights the effectiveness of AGN feedback, and the delayed resumption of star formation that this entails.

Gas particles are labelled as `star forming' if they have an instantaneous SFR > 0; this SFR is pressure and metallicity dependent, and determines the probability that a gas particle will be converted in to a star particle.
Since \eagle\ does not resolve individual molecular clouds, the mass of this star-forming gas can be thought of as a proxy for the cold gas mass.
The bottom panel of \fig{gas_fraction} shows the star-forming gas fraction over the total mass of gas in the halo.
The behaviour in this space is more complicated with halo mass, rising from lower values in low mass haloes to a peak at $10^{11.5} \mathrm{M_{\odot}}$, before falling off again in higher mass haloes.
This behaviour broadly fits with estimates of star formation efficiency as a function of halo mass derived from abundance matching techniques \citep{behroozi_average_2013}, including the position of the peak at $z = 5$.

Our passive galaxy selection again has lower star forming gas fractions than the overall population.
Combined with the lower gas fractions seen in the top panel of \fig{gas_fraction}, this suggests that not only is the reservoir of gas available for star formation depleted after expulsion by AGN feedback, the gas that remains is also heated by AGN feedback, preventing it from becoming star forming.

\subsubsection{Stellar Age and Formation Time}
\label{sec:age_formation}

\begin{figure*}
    \centering
    \includegraphics[width=\textwidth]{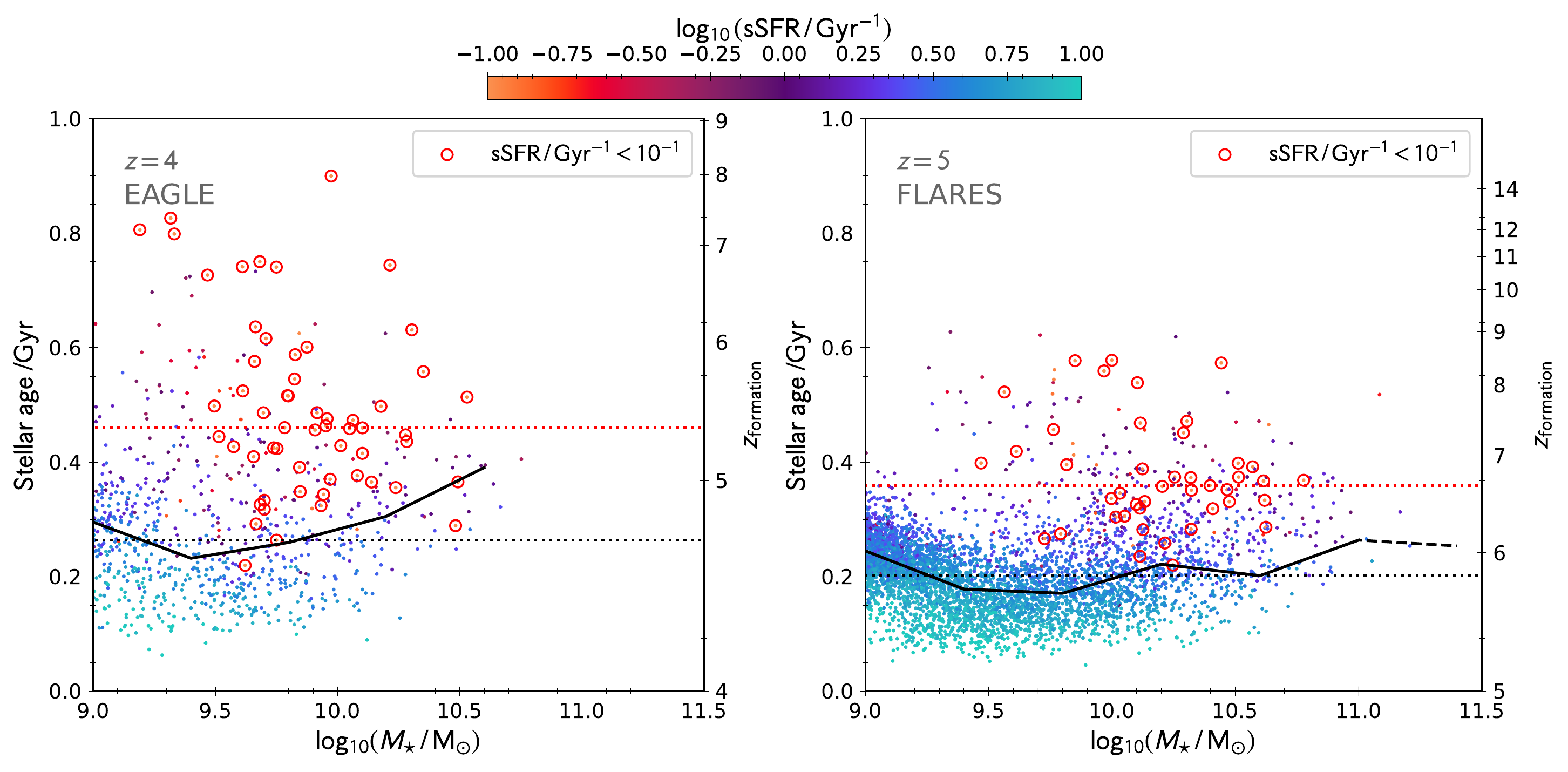}
    \caption{
        Stellar formation time against stellar mass, coloured by sSFR.
        Passive galaxies ($\mathrm{sSFR < 10^{-1} \,/\, Gyr^{-1}}$) are highlighted with red circles.
        The horizontal black and red lines show the median formation time for the whole population and the passive population, respectively, (where $M_{\star} \,/\, \mathrm{M_{\odot}} > 10^9$).
        \textit{Left:} all \eagle\ galaxies at $z = 4$, with the binned median in black.
        \textit{Right:} all \flares\ galaxies at $z = 5$, with the binned \textit{weighted} median of all regions combined shown as the black line (dashed where there are fewer than ten galaxies in a bin).
    }
    \label{fig:formation_time}
\end{figure*}

\fig{formation_time} shows the stellar age of each galaxy, defined as the mean of the initial mass weighted stellar age of each star particle within the aperture (see \sec{aperture_timescale}), in \eagle\ and \flares\ at $z = 4$ and $5$, respectively.
The redshift corresponding to this stellar age at the given redshift, defined as $z_{\mathrm{formation}}$, is also shown for reference.
There is a weak trend with stellar mass at both redshifts, where galaxies with $M_{\star} \,/\, \mathrm{M_{\odot}} \sim 10^{9.5}$ have the youngest stellar ages, and lower / higher mass galaxies are, on average, older.
There is also a trend in formation with sSFR at fixed stellar mass, whereby older galaxies tend to have lower star formation rates.
This is not so surprising at fixed stellar mass; in order to reach the same stellar mass, those galaxies with lower sSFRs must necessarily have formed their stars earlier.
Indeed, when we isolate the quiescent population, we see that they have some of the oldest stellar ages.
In \flares\ at $z = 5$, the median age for all galaxies with stellar masses $M_{\star} \,/\, \mathrm{M_{\odot}} > 10^9$ is $214^{+62}_{-67}$ Myr, whereas for passive galaxies this rises to $364^{+110}_{-74}$ Myr.
In \eagle\ at $z = 4$ the median age for all galaxies is $277^{+134}_{-100}$ Myr, and $464^{+179}_{-113}$ Myr for passive galaxies.

\begin{figure}
    \centering
    \includegraphics[width=\columnwidth]{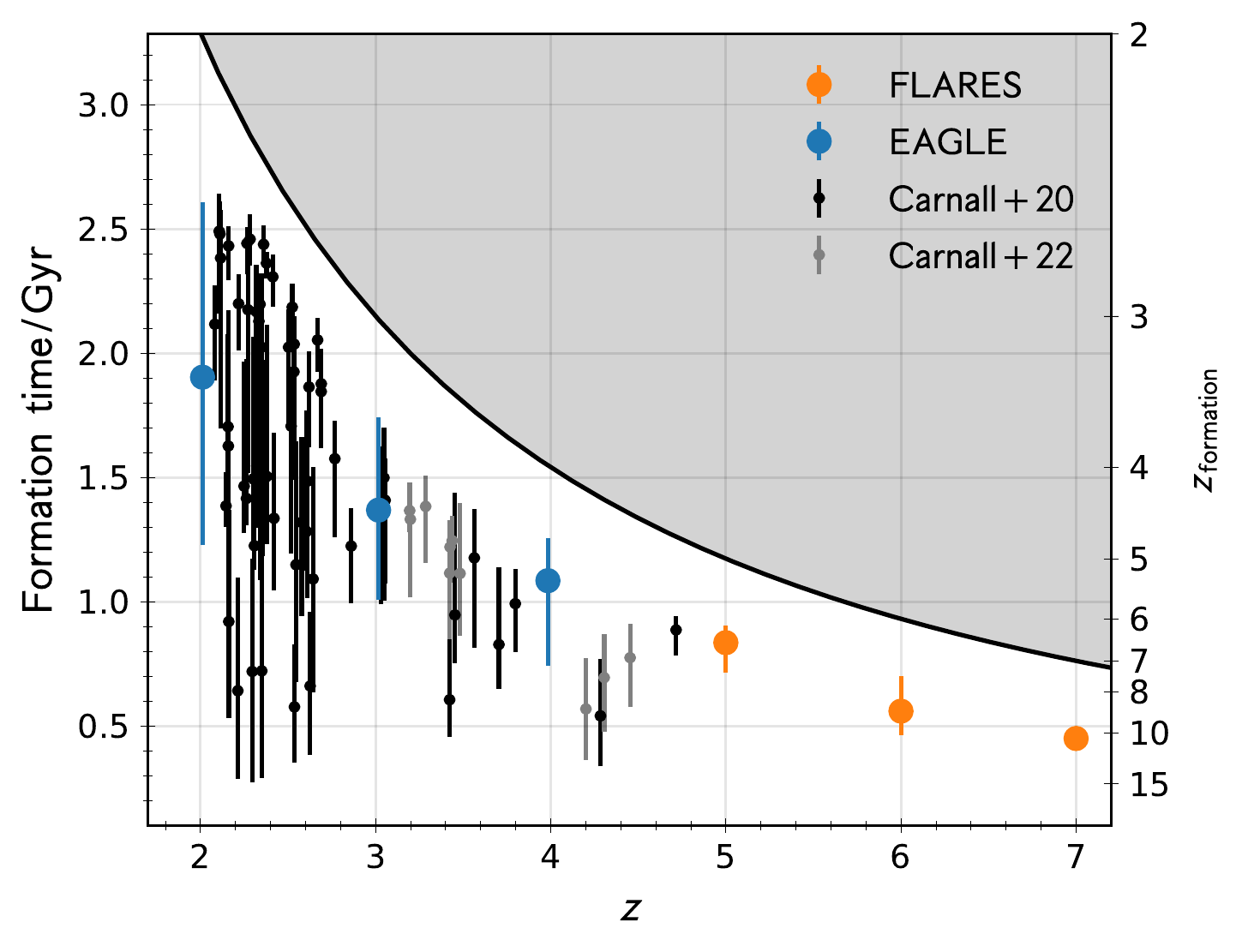}
    \caption{
        Redshift evolution of the stellar formation time of passive galaxies ($\mathrm{sSFR < 10^{-1} \,/\, Gyr^{-1}}$, $M_{\star} \,/\, \mathrm{M_{\odot}} > 10^9$).
        The median and $5^{\mathrm{th}}-95^{\mathrm{th}}$ percentiles of galaxies in \flares\ (orange) and \eagle\ (blue) at discrete redshifts between $2 < z < 7$ are shown.
        Observational estimates of individual `robust' candidates from \protect\cite{carnall_timing_2020} and \protect\cite{carnall_surprising_2023} are shown in black and grey, respectively.
        The black shaded region shows the range of values ruled out by the age of the Universe.
    }
    \label{fig:form_time_redshift}
\end{figure}

\fig{form_time_redshift} shows the stellar age again, but expressed as the \textit{formation time}, or the age of the universe at the given stellar age.
Passive galaxies (defined as $\mathrm{sSFR < 10^{-1} \,/\, Gyr^{-1}}$; $M_{\star} \,/\, \mathrm{M_{\odot}} > 10^9$) at $z \geqslant 5$ tend to have formed within the first billion years of the universe's history, evolving to later formation times at lower redshifts.
These results are in qualitative agreement with inferred ages of passive galaxies in HST and JWST data from \cite{carnall_timing_2020,carnall_surprising_2023}.
However, there are a number of sources at $4 < z < 5$ in these observational samples with inferred formation redshifts $z_{\mathrm{formation}} \sim 10$, making them potentially the descendants of the massive galaxies detected at these high redshifts in \cite{labbe_population_2023}.
In \flares\ and \eagle\ we do not see passive galaxies with such early formation times until $z \geqslant 6$.
At lower redshifts the median formation times in \eagle\ are in good agreement with the range of values inferred in \cite{carnall_timing_2020}.

\subsubsection{Quenching Timescales}

\begin{figure}
    \centering
    \includegraphics[width=\columnwidth]{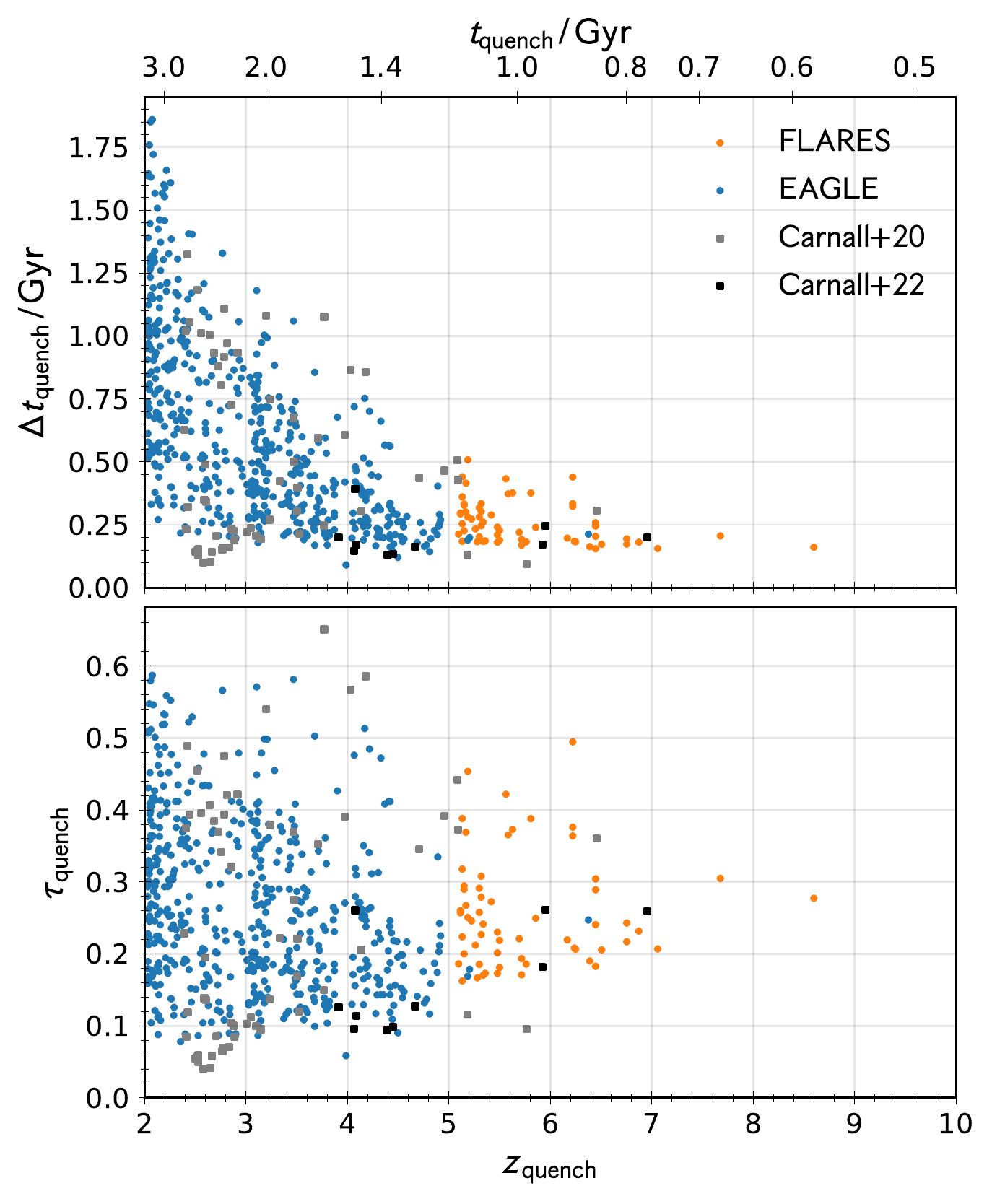}
    \caption{
        Quenching timescales for passive galaxies ($\mathrm{sSFR < 10^{-1} \,/\, Gyr^{-1}}$) in \flares\ (orange) and \eagle\ (blue) between $2 < z < 7$ against the quenching redshift / age ($z_{\mathrm{quench}}$ / $t_{\mathrm{quench}}$).
        Observational estimates for individual galaxies from \protect\cite{carnall_timing_2020,carnall_surprising_2023} are shown in grey and black, respectively.
        \textit{Top:} quenching timescale, $t_{\mathrm{quench}}$.
        \textit{Bottom:} normalised quenching timescale, $\tau_{\mathrm{quench}}$.
    }
    \label{fig:quenching_timescales}
\end{figure}

The star formation histories shown in \fig{SFH} and \ref{fig:SFH_B} exhibit significant diversity, particularly in the time it takes for each galaxy to quench.
In order to quantify this we use the quenching timescale \citep{tacchella_fast_2022,carnall_inferring_2018},
\begin{align}
    \Delta t_{\mathrm{quench}} = t_{\mathrm{quench}} - t_{\mathrm{form}} \;\;,
\end{align}
where $t_{\mathrm{quench}}$ is the age of the universe at the point when the galaxy quenched, and $t_{\mathrm{form}}$ is the formation time.
$t_{\mathrm{quench}}$ is defined, using the full star formation history, as the time at which the sSFR last fell below $10^{-1} \,/\, \mathrm{Gyr^{-1}}$.
The top panel of \fig{quenching_timescales} shows $\Delta t_{\mathrm{quench}}$ for passive galaxies in \flares\ and \eagle\ against $z_{\mathrm{quench}}$ (identical to $t_{\mathrm{quench}}$, but expressing the redshift).
Galaxies tend to quench very quickly at early times ($z > 6$), with $\Delta t_{\mathrm{quench}}$ < 200 Myr.
At later times, the quenching timescales increase dramatically, reaching up to 1 Gyr for some galaxies by $z = 2$.
The redshift evolution is in qualitative agreement with observationally estimated quenching timescales in \cite{carnall_timing_2020,carnall_surprising_2023}, and the quantitative values are in very good agreement across the entire redshift range probed.

It is also useful to show the normalised quenching timescale, since galaxies' dynamical timescales evolve significantly with redshift.
We define this using the age of the universe at the point the galaxy quenches, $t_{\mathrm{quench}}$,
\begin{align}
    \tau_{\mathrm{quench}} = \frac{\Delta t_{\mathrm{quench}}}{t_{\mathrm{quench}}} \;\;.
\end{align}
The bottom panel of \fig{quenching_timescales} shows that there is little redshift evolution in the median of this relation between $2 < z < 7$, however some galaxies at later times have much higher normalised quenching timescales.
This is, again, in quantitative agreement with individual galaxies and the overall redshift evolution estimated in \cite{carnall_timing_2020,carnall_surprising_2023}.

We caution that a fairer comparison between the simulations and the observations would implement a similar observational selection, and that therefore any comparisons between the two should be made with caution.
We highlight some of the issues with typical passive selection criteria in \sec{obs}.

\subsection{Discussion}
The analysis in this section paints a compelling picture of the cause of passivity in high-$z$ galaxies in \flares\ and \eagle.
The analysis of individual star formation and black hole accretion histories suggests a strong anti--correlation between the two.
And when looking at the black hole stellar mass relation we see that passive galaxies lie above the median relation at $z = 5$; their black holes are over--massive at fixed stellar mass.
Since black hole mass is roughly equivalent to the total feedback energy injected into the ISM of a galaxy, particularly at high-z where black hole mergers are rare, these passive galaxies have experienced more feedback energy per unit stellar mass.
We also see that, when looking at the host halo as a whole, total gas and star forming gas fractions are lower in passive galaxies, both symptoms of strong black hole feedback heating and evacuating gas from the centrals regions of galaxies.

Stellar feedback does not seem to play a large part in causing passivity in galaxies of any mass, at least above our resolution limit ($M_{\star} \,/\, \mathrm{M_{\odot}} > 10^{8}$).
This is shown by the prevalence of rising star formation histories in the overall galaxy population at $z \geqslant 5$ \citep{wilkins_first_2022}; if stellar feedback was an important contributor, we should see evidence for stronger self-regulation as star formation rates rise into the hundreds of solar masses regime.
However, this may be a result of the thermal feedback scheme from stellar sources in \eagle; previous results have shown that a combination of thermal and kinetic feedback from stars with a suitable algorithmic implementation can be effective in removing gas from galactic halos \citep{merlin_formation_2012}.
This combination avoids overcooling, if the SFR is high enough.

\begin{figure}
    \centering
    \includegraphics[width=20pc]{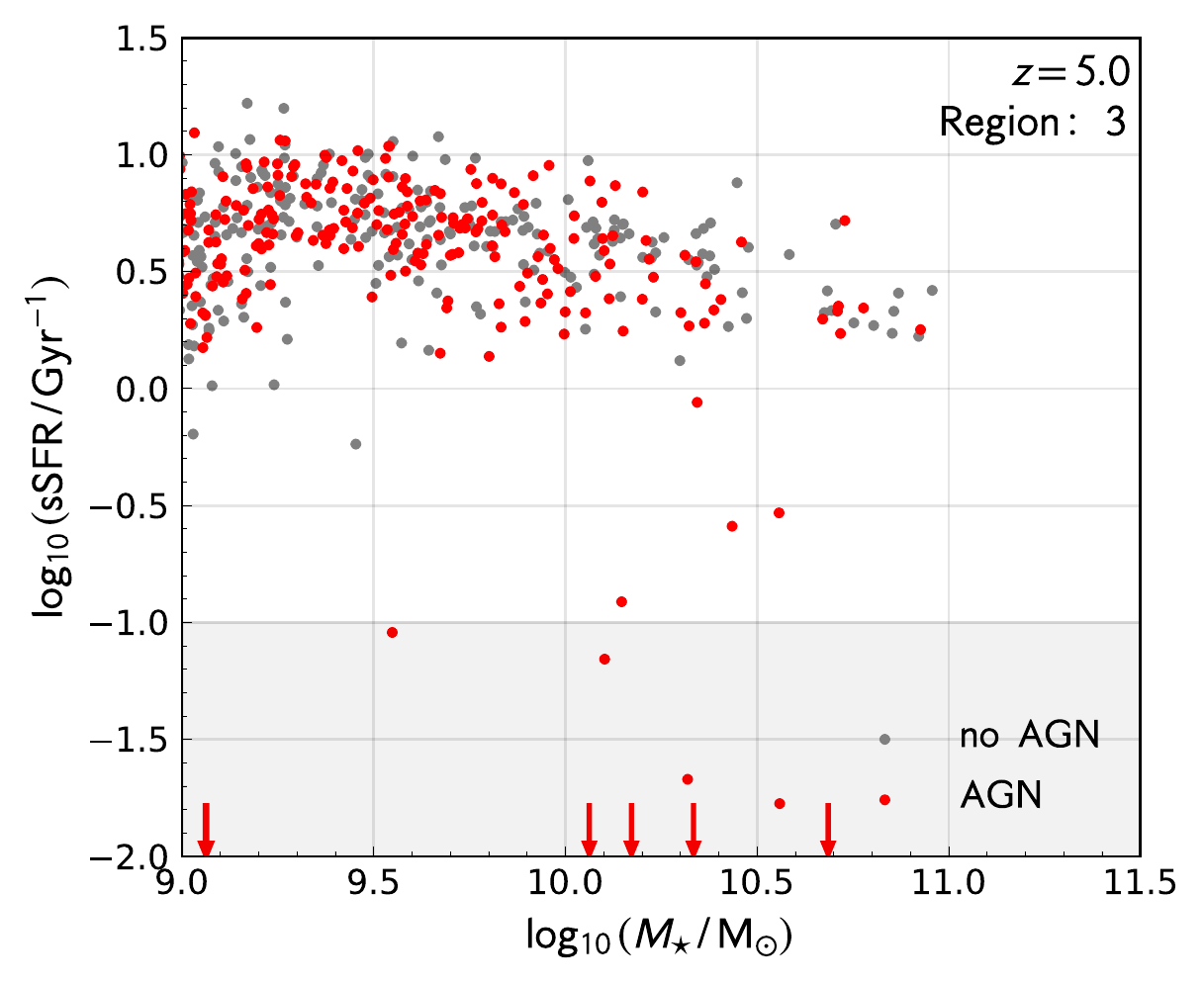}
    \caption{
      The star forming sequence in \flares\ at $z = 5$ in one of the highly overdense simulated regions.
      Star formation rates and stellar masses are measured within a 5 kpc aperture.
      We show individual galaxies in this region for simulations with and without the effect of AGN turned on (red and grey points, respectively).
      Galaxies with sSFR's below the plot limits are shown with red arrows.
      Turning off AGN leads to no passive galaxies ($\mathrm{sSFR < 10^{-1} \,/\, Gyr^{-1}}$) being produced in this region at $z = 5$.
    }
    \label{fig:noAGN_SFS}
\end{figure}

Using merger graphs constructed by the MErger Graph Algorithm \citep[\textsc{Mega}][]{roper_mega_2020} we have also investigated the effect of major mergers\footnote{We define major mergers as mergers where the lower mass galaxy of the merging pair contributes at least 10\% to the final total mass}, and whether they also contribute to the passive population.
We analysed galaxies in the $z = 5$ snapshot and searched for those with major mergers between the selection redshift and the previous snapshot, at $z = 6$.
We find that, in the overall galaxy population, 15\% of galaxies undergo major mergers \citep[in reasonable agreement with extrapolations from lower redshift observational measurements;][]{duncan_observational_2019}, whereas in the passive population this is just 7\%.
This suggests that major mergers are not the dominant cause of passivity in our passive selection, and does not lead to passivity in the majority of galaxies that experience major mergers.

As a final conclusive test, we ran one of our regions, `03', with AGN feedback turned off.
This region is significantly overdense, and was chosen as it contains 9 passive galaxies in the shown mass range (in the fiducial run).
\fig{noAGN_SFS} shows galaxies in this region on the star forming sequence.
It is clear that in the no-AGN run there are no passive galaxies; the lowest sSFR's in the no-AGN run are $> -0.3 \,/\, \mathrm{Gyr^{-1}}$.

\begin{figure*}
    \centering
    \includegraphics[width=\textwidth]{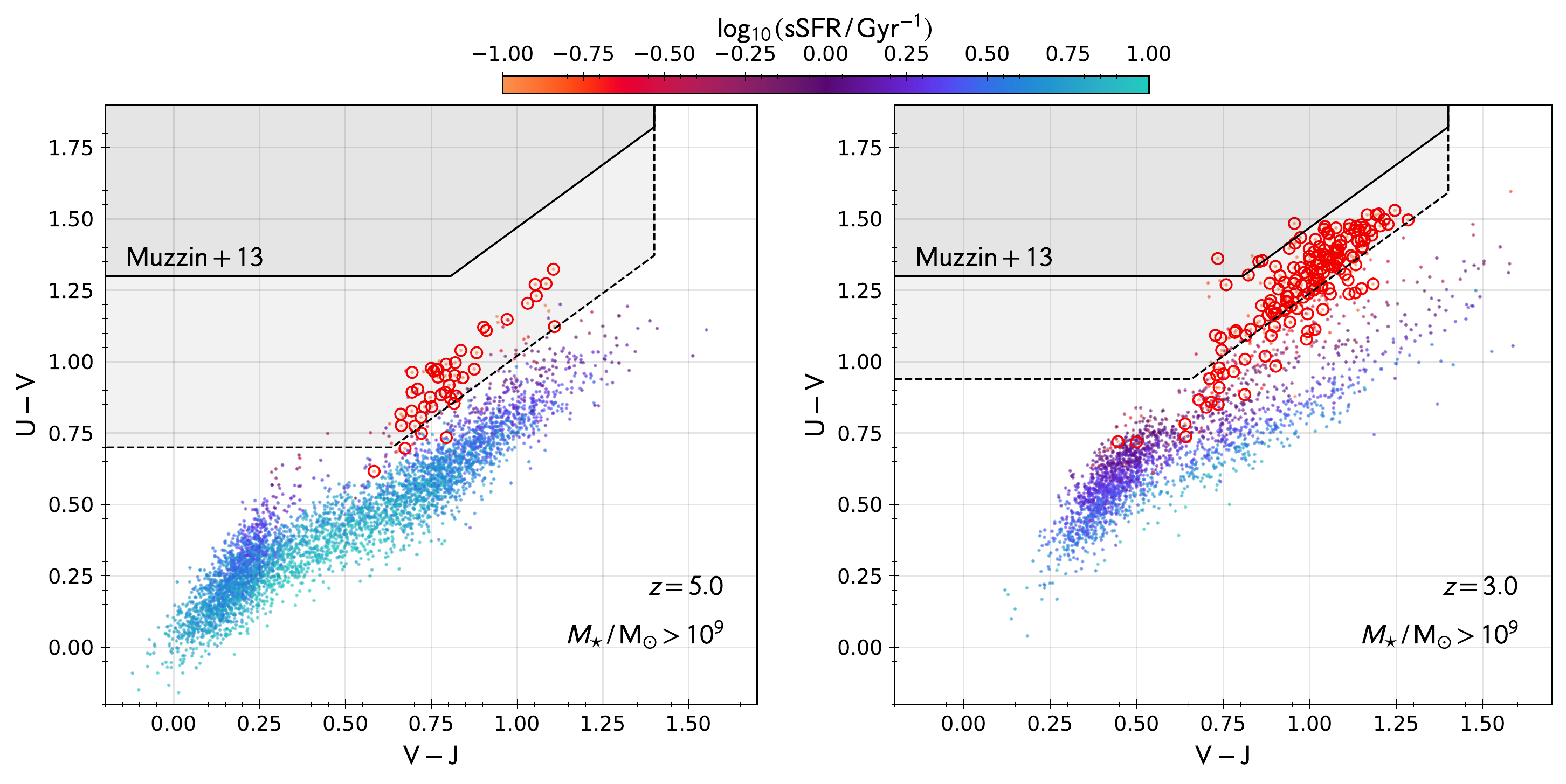}
    \caption{
        Rest-frame UVJ diagram for galaxies with $M_{\star} \,/\, \mathrm{M_{\odot}} > 10^{9}$, coloured by sSFR.
        Passive galaxies ($\mathrm{sSFR < 10^{-1} \,/\, Gyr^{-1}}$) are circled in red.
        Colours are calculated using the BPASS model \protect\citep{stanway_re-evaluating_2018}.
        The typical passive selection space \protect\citep{muzzin_evolution_2013} is shown in dark grey with solid black border.
        The new selection region presented in \sec{UVJ_selection}, incorporating redshift evolution, and extending to bluer UV colour, is shown in lighter grey with a black dashed border.
        \textit{Left panel:} all \flares\ galaxies at $z = 5$.
        \textit{Right panel:} all \eagle\ galaxies at $z = 3$.
    }
    \label{fig:selection.UVJ}
\end{figure*}

Another compelling aspect of the analysis presented is the formation times and quenching timescales of these passive galaxies.
Almost by definition, passive galaxies are expected to have formed earlier than their star forming counterparts (at fixed stellar mass), however it is interesting to understand quantitatively how old they are, as this may affect their observable colours.
The analysis of individual star formation histories shows a handful of galaxies that become quiescent at $z > 7$, and remain approximately so up to $z = 5$, a duration of $\sim 400$ Myr.
This leads to formation redshifts of $z > 8$ for a number of galaxies.
These formation times are in good agreement with recent constraints from \cite{carnall_surprising_2023}, which suggests this might be an analogous population of massive galaxies forming at very early times.

\section{Observational Measures of Quiescence}
\label{sec:obs}

So far we have looked at passive galaxies in \flares\ and \eagle\ selected through their intrinsic sSFR.
However, in many observational studies passive galaxies are first identified through their colours, typically in rest--frame UVJ space \citep[\textit{e.g.}][]{muzzin_evolution_2013,straatman_substantial_2014,schreiber_near_2018}.
In this section we forward model the UV-NIR emission from galaxies in \flares\ at a range of redshifts, and evaluate the UVJ selection, as well as any intrinsic biases introduced by our modelling assumptions.
We also evaluate the selection of passive galaxies using observer--frame JWST colours from the NIRCam and MIRI instruments.

\subsection{Forward Modelling Galaxy Emission}
\label{sec:forward_modelling}

In order to predict the emission from our modelled galaxies we implement a forward modelling pipeline first introduced in \cite{vijayan_first_2021}.
Each stellar particle is treated as a simple stellar population (SSP), where all the stars are coeval with identical metallicity.
These SSPs are associated, based on their age and metallicity, with a single SPS model; for our fiducial model we use BPASS \citep{stanway_re-evaluating_2018}, but explore other models later in this section.
For star particles with ages less than 10 Myr we assume the stars are still resident within their birth clouds, and therefore subject to birth cloud attenuation of their Lyman continuum emission, which is reprocessed into nebular emission.
To calculate this emission we use the approach demonstrated in \cite{wilkins_nebular-line_2020}, whereby the emission from these young SSPs is used as input to the CLOUDY photoionisation code \citep{ferland_2017_2017}.
The metallicity of the cloud is assumed to be identical to that of the parent star particle, and abundances and dust depletion factors are adopted from \cite{gutkin_modelling_2016}.
We vary the ionisation parameter based on the input ionising spectrum, using a fixed reference ionisation parameter of $\mathrm{log_{10}} U = -2$ at $t = 1 \mathrm{Myr}$ and $Z = 0.02$.

Dust attenuation is treated using a line of sight model, whereby the metal column density along the $z$-direction is calculated towards each star particle, assuming the SPH kernel of each gas particle.
This allows for differential attenuation of different stellar populations, based on the full star--dust geometry.
The dust-to-metals ratio for each galaxy is calculated using the fitting function presented in \cite{vijayan_detailed_2019}, which is a function of the gas-phase metallicity and mass--weighted stellar age of all particles.

Further details on various steps in this pipeline are provided in \cite{wilkins_nebular-line_2020,vijayan_first_2021}.
The only major modification to the pipeline presented in these works is the aperture within which we calculate the luminosities and fluxes.
Here, we assume the same 3D aperture used to calculate physical properties (see \sec{aperture_timescale}), with radius closest to $R = 4 \times R_{1/2,\star}.$\footnote{for the following discrete apertures: $R$ = [1, 3, 5, 10, 20, 30, 40, 50, 70, 100] $\mathrm{pkpc}$}
The effect of this choice on the measured UVJ colours is shown in \app{aperture_UVJ}, but in general this aperture choice acts to make galaxy colours \textit{redder}, since residual star formation on the outskirts of galaxies is excluded from the aperture.

\subsection{Rest-frame UVJ Selection}
\label{sec:UVJ_selection}

\fig{selection.UVJ} shows all \flares\ galaxies at $z = 5$ on the rest-frame UVJ colour-colour diagram, coloured by sSFR, assuming our fiducial forward model.
We also show the typical UVJ selection space for passive galaxies at $1 < z < 4$ from \cite{muzzin_evolution_2013}.
What is immediately obvious is that no \flares\ galaxies populate this passive selection space, whether they are passive or not.
Almost all galaxies have VJ colour that lies within the passive selection range, but the predicted UV colours of passive galaxies are too blue by at least 0.2 magnitudes.
\fig{selection.UVJ} also shows the UVJ diagram at $z = 3$ for \eagle.
A number of passive galaxies now populate the \cite{muzzin_evolution_2013} selection space, however the majority of our passive selection does not.
This is in agreement with the results of \cite{schreiber_near_2018}, who found that UVJ selection at these redshifts was pure, but incomplete.

\begin{figure}
    \centering
    \includegraphics[width=20pc]{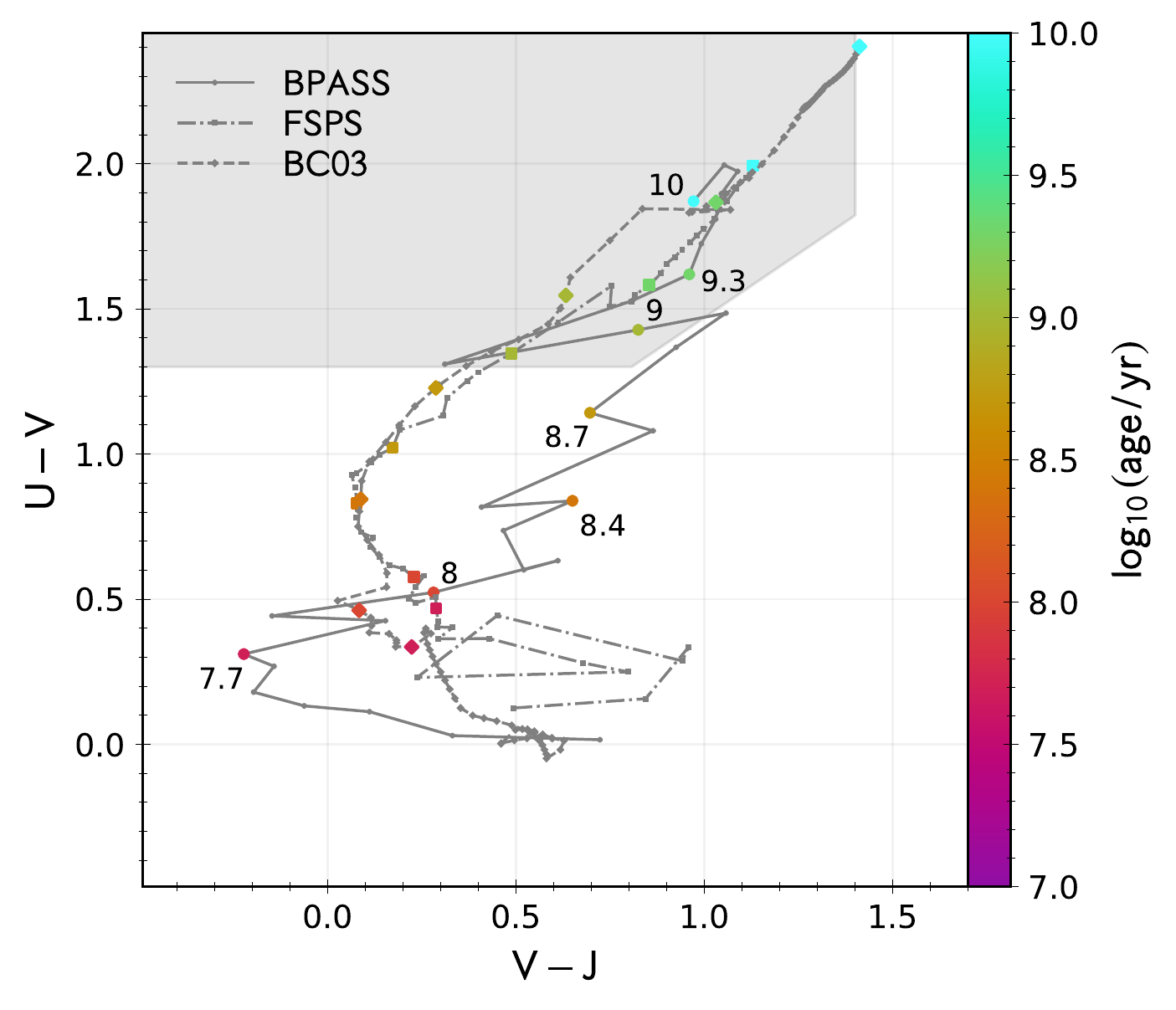}
    \caption{
        Tracks in rest-frame UVJ space for simple stellar populations ($Z = 0.01$), assuming different SPS models (BC03, FSPS \& BPASS; dashed, dashed-dotted and solid lines, respectively).
        Coloured points show where each population reaches a given age.
        The typical passive selection space \protect\citep{muzzin_evolution_2013} is shown in dark grey.
    }
    \label{fig:UVJ_tracks}
\end{figure}

To further investigate, we first explored a number of other stellar population synthesis (SPS) models.
FSPS \citep{conroy_propagation_2009,conroy_propagation_2010} and BC03 \citep{bruzual_stellar_2003} both do not take into account the effect of binary stellar evolution on the integrated emission from a given stellar population.
Binary stellar evolution can have a number of effects; for example, interactions between stars in binary or higher multiple stellar systems typically lead to hotter stellar surface temperatures, which can increase the hardness of the radiation emitted.
\fig{UVJ_tracks} shows tracks in UVJ space for BPASS, FSPS and BC03 for a simple stellar population with fixed metallicity ($Z = 0.01$) that evolves secularly through time.
It is clear that at lower ages, less than a billion years, both BC03 and FSPS predict very different behaviour in colour space to BPASS, which shows much bluer VJ colours, but for older ages the models generally converge.
The key insight, however, is that for populations with ages less than 500 Myr none of the models predict colours within the passive selection space of \cite{muzzin_evolution_2013}.
\cite{merlin_chasing_2018} found similar behaviour - in order to reach the passive selection space at these high redshifts requires very high formation ages (close to the age of the Universe at $z \geqslant 5$) and strongly truncated SFHs.

\begin{figure}
    \centering
    \includegraphics[width=20pc]{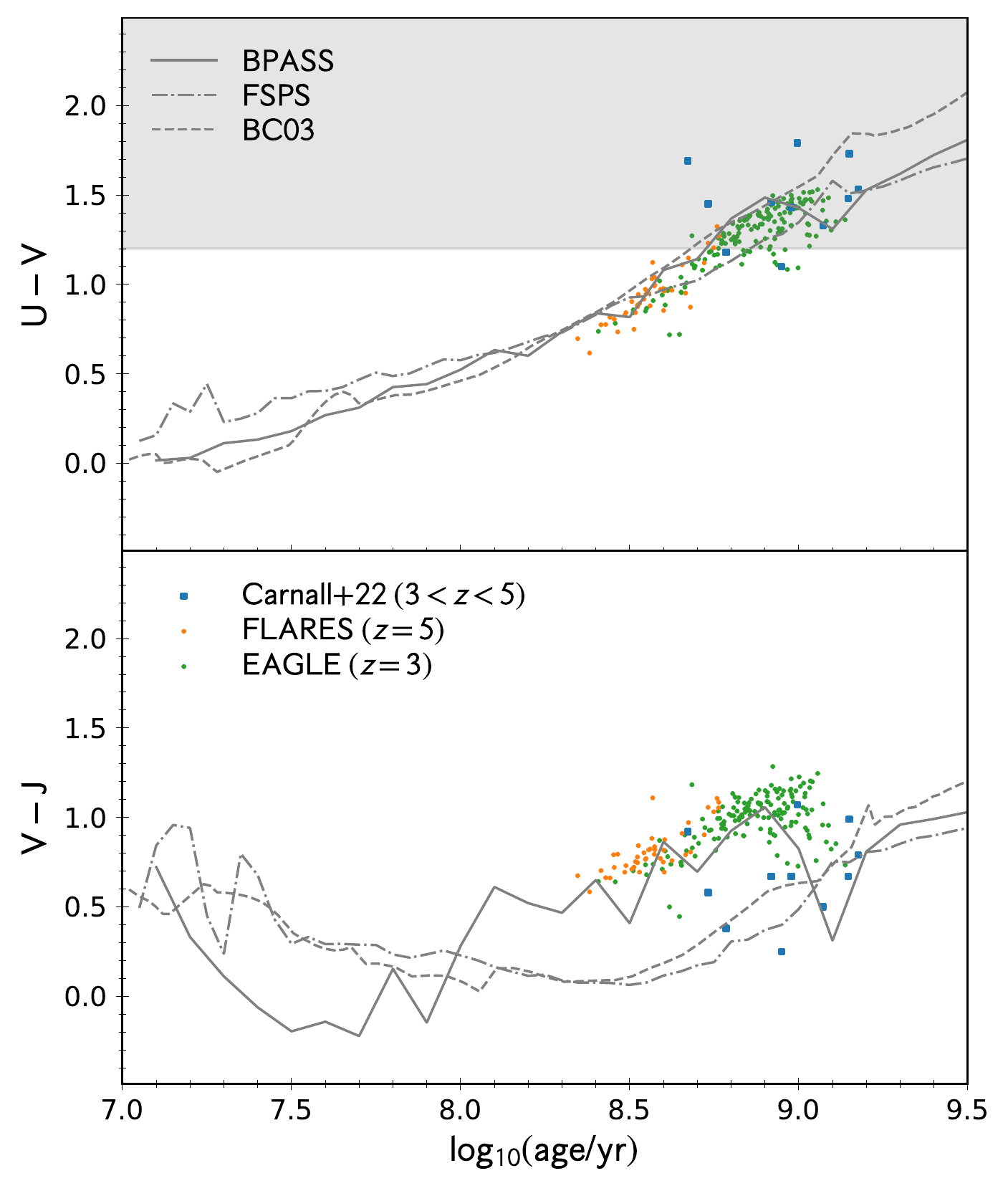}
    \caption{
        Evolution of rest-frame UV and VJ colours assuming different SPS models (BC03, FSPS, BPASS).
        The grey region in the top panel is the minimum UV colour required for inclusion in the \protect\cite{muzzin_evolution_2013} passive selection space.
        We also show passive \flares\ galaxies at $z = 5$ (orange points) and the recent $3 \leqslant z \leqslant 5$ candidates from \protect\cite{carnall_surprising_2023}, where the $x$-axis shows their (initial mass--weighted) formation time.
    }
    \label{fig:UVJ_evolution}
\end{figure}

To show this more explicitly, \fig{UVJ_evolution} shows the UV and VJ colours individually as a function of stellar population age, for each SPS model.
Here it is clear that the main difference between the models is in VJ space, whereas in UV there is a strong positive correlation between the age and colour, regardless of SPS model.
We also show \flares\ and \eagle\ galaxies at $z = 5$ and $3$, respectively, and the recent results from \cite{carnall_surprising_2023}, where instead of SPS model age we show the initial mass--weighted formation time.
From this we can see that, in \flares, due to the higher redshift selection, the formation times are lower ($< 600 \, \mathrm{Myr}$), and therefore their UV colours do not reach the passive space.
Conversely, the candidates in \cite{carnall_surprising_2023}, due to the slightly lower selection redshift, reach formation ages of $\sim 1 \, \mathrm{Gyr}$, sufficient to move them into the passive selection space, at least in UV.
\cite{akins_quenching_2022} found that a similar timescale is necessary for inclusion in the passive selection space when looking at the \textsc{Simba} simulations.

An interesting aside is the evolution in VJ space.
Here, \flares\ and \eagle\ galaxies follow the BPASS tracks as expected, since this model was used in the forward modelling pipeline.
In \eagle, the slightly higher ages means that they are sufficiently red in UV to reach the passive selection region, however \fig{selection.UVJ} shows how their VJ colours are still too red - this appears to be due to the sensitivity of VJ colour to SPS model choice, with BPASS predicting redder colour than BC03 and FSPS for populations with age $\sim 1 \; \mathrm{Gyr}$.
The \cite{carnall_surprising_2023} results do not seem to prefer any of the models, despite assuming \cite{bruzual_stellar_2003} in the SED fitting procedure.
One reason may be that stellar birth metallicity introduces further scatter in the UVJ colours.

\fig{selection.UVJ} shows a new, more liberal selection region, fitted to our \flares\ and \eagle\ data at $z \leqslant 5$, that includes an explicit dependence on redshift, defined as
\begin{align}
    V \,-\, J &< 1.4 \\
    U \,-\, V &> 1.42 - 0.12 \times (1+z) \\
    U \,-\, V &> (V \,-\, J) \times 0.88 - 0.11 \times (1+z) + 0.8 \,\,.
\end{align}
This region is identical to \cite{muzzin_evolution_2013} at $z = 0$, but selects bluer objects at higher redshifts.
There is significant overlap of passive and star-forming objects in UVJ space at these redshifts, however this parametrisation maximises the completeness and purity of our simulated sample within this redshift range.

\begin{figure}
    \centering
    \includegraphics[width=20pc]{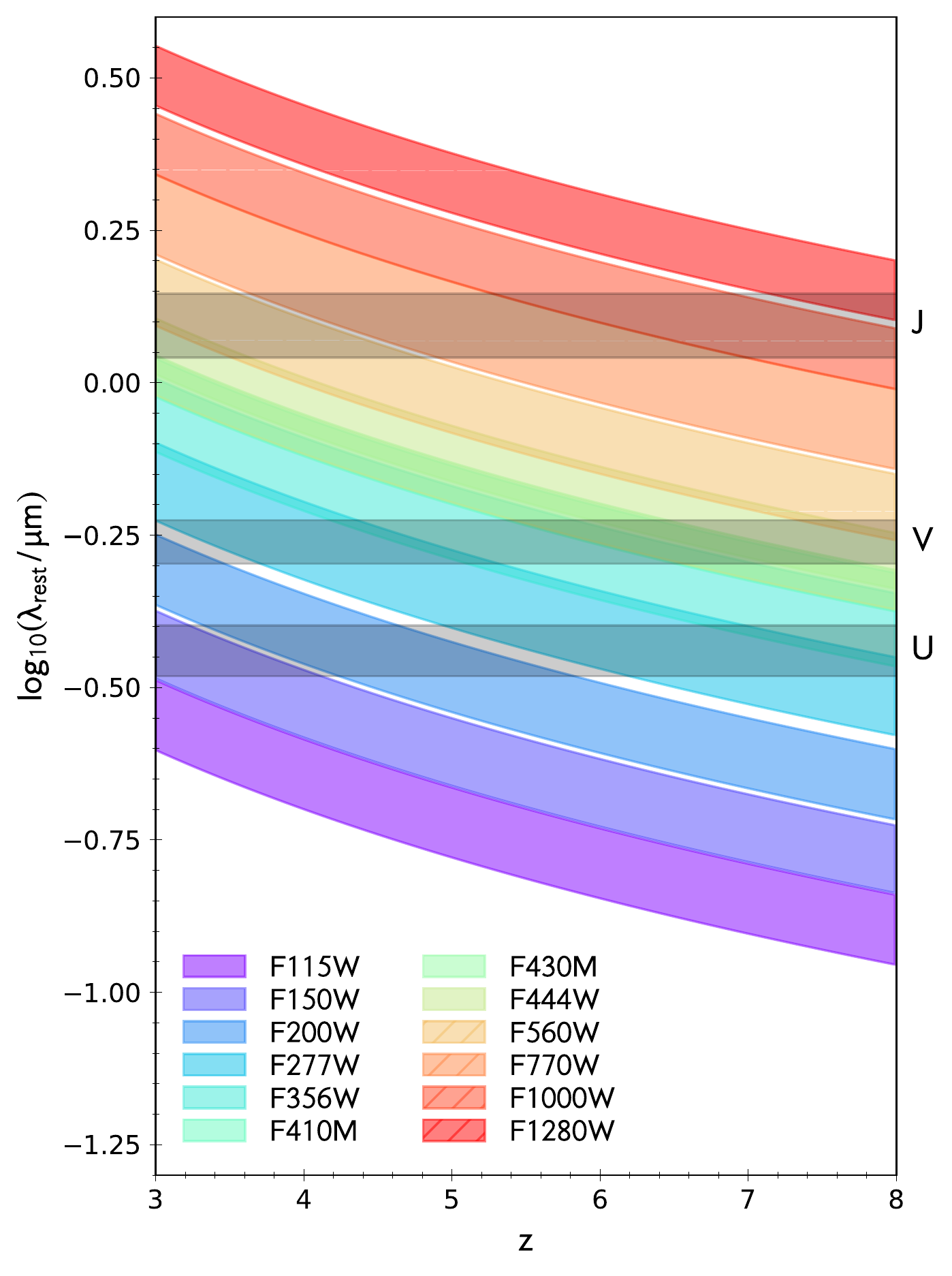}
    \caption{
        Evolution of rest--frame wavelength of JWST NIRCam and MIRI filters as a function of observed redshift.
        The rest--frame wavelength of the U, V and J filters are shown by the grey bands.
        The J filter is only probed at $z \leqslant 3$ with NIRCam; above this redshift only UV colour is probed in a model-independent way with NIRCam.
    }
    \label{fig:jwst_uvj_evolution}
\end{figure}

In summary, the typical rest--frame selection using UVJ struggles to select passive galaxies where the underlying stellar population is younger than 1 Gyr.
These galaxies are passive, in the sense that their sSFRs, as measured using their SFRs averaged over the past 50 Myrs, are below our passive selection criteria.
However, their `evolved' stellar populations are still very young, precluding their inclusion in the typical UVJ colour selection space due to their intrinsically blue UV colours; passive galaxies in \flares\ are relatively young, analogous to rapidly quenched galaxies in the later Universe \citep{park_rapid_2022}.
A number of observational studies have found, through SED fitting, similar behaviour at high redshift, whereby galaxies with low sSFR estimates lie outside the typical UVJ selection space \citep{merlin_chasing_2018,schreiber_near_2018}.
In \cite{merlin_chasing_2018} they attribute this to uncertainties and assumption in SED modelling, particularly the form of the star formation history adopted, which agrees with our findings.
New selection regions in UVJ, such as those presented here, or new rest--frame colour spaces \citep[\textit{e.g.}][]{antwi-danso_beyond_2023,long_efficient_2023}, are required for detecting passive objects at $z \geqslant 5$.

\begin{figure*}
    \centering
    \includegraphics[width=\textwidth]{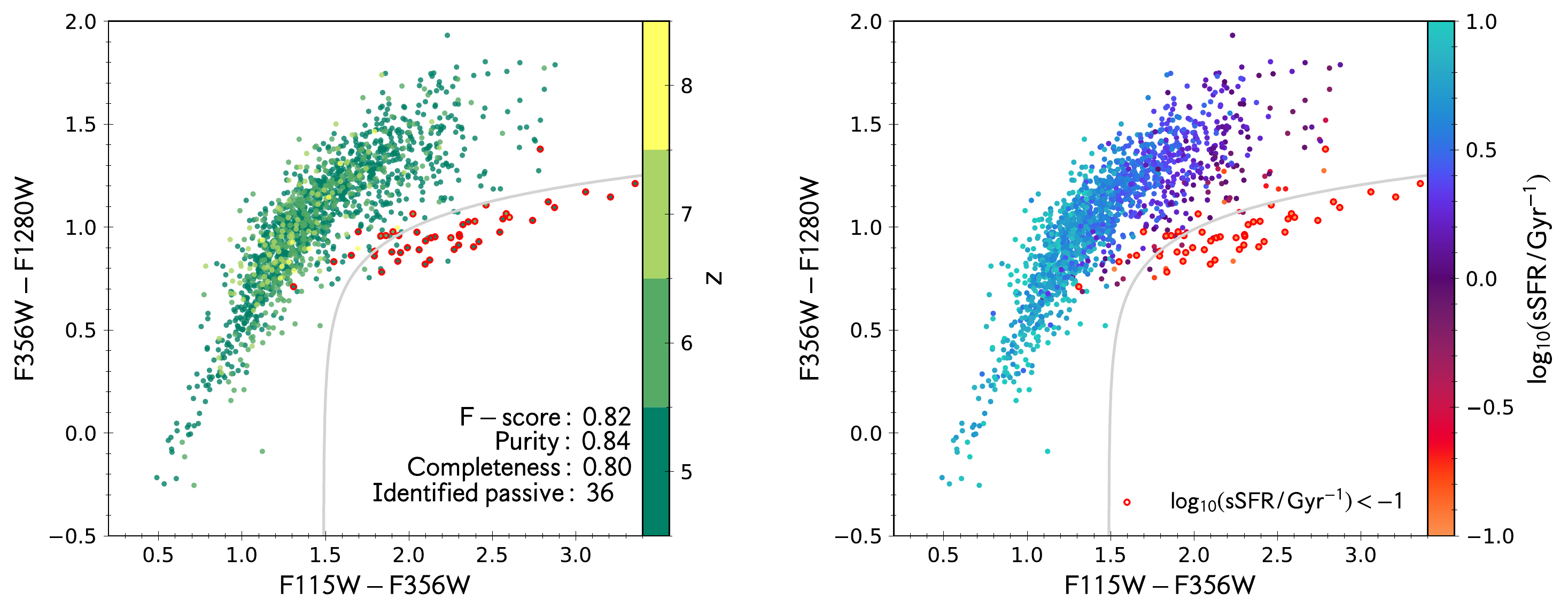}
    \includegraphics[width=\textwidth]{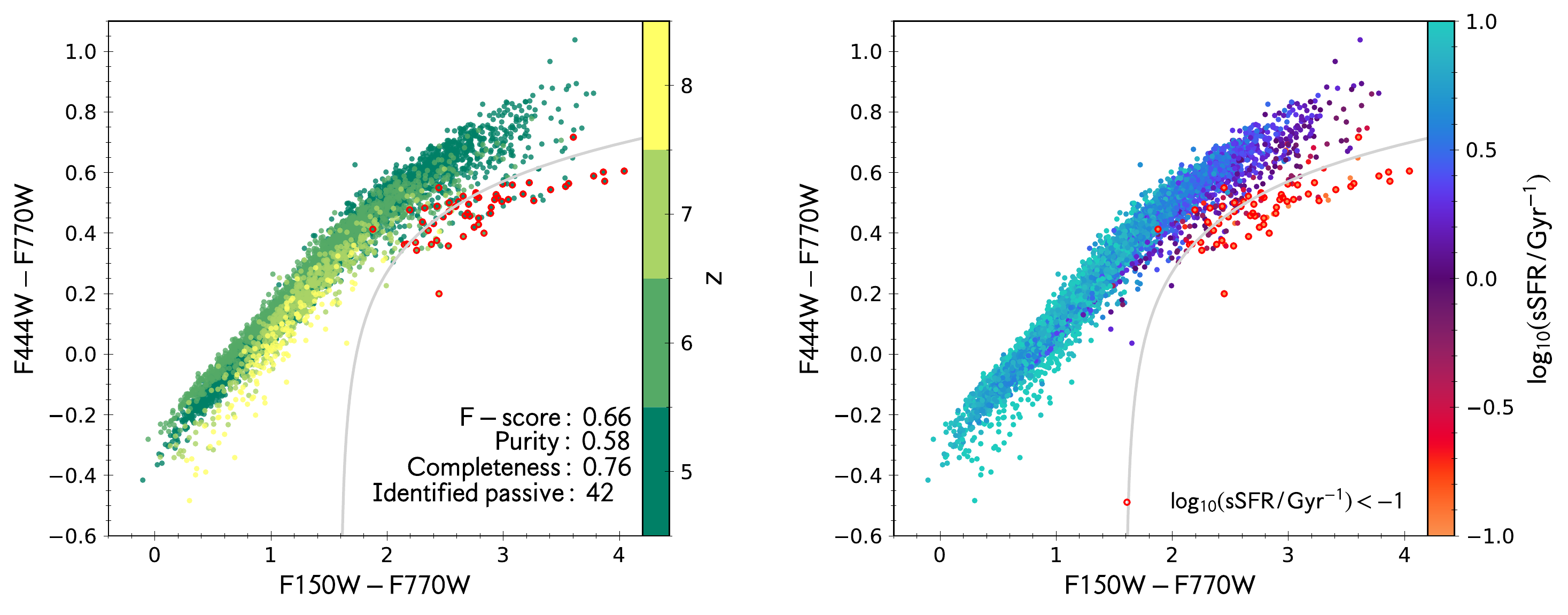}
    \caption{
        Colour -- colour plots of all \flares\ galaxies between $5 \leqslant z \leqslant 8$ for different combinations of NIRCam and MIRI bands.
        In the first column points are coloured by the redshift of the snapshot, and in the second column by their sSFR.
        Passive galaxies ($\mathrm{sSFR < 10^{-1} \,/\, Gyr^{-1}}$) are highlighted with red circles.
        Grey curves show the chosen passive galaxy selection space.
        The purity and completeness of this selection, as well as the number of selected passive galaxies, is quoted in each panel.
        \textit{Top row:} the $\mathrm{[F115W - F356W]}$ \& $\mathrm{[F356W - F1280W]}$ space.
        \textit{Bottom row:} the $\mathrm{[F150W - F770W]}$ \& $\mathrm{[F444W - F770W]}$ space.
    }
    \label{fig:F115W_F356W_F356W_F1280W_ssfr_curve}
\end{figure*}

\begin{figure*}
    \centering
    \includegraphics[width=\textwidth]{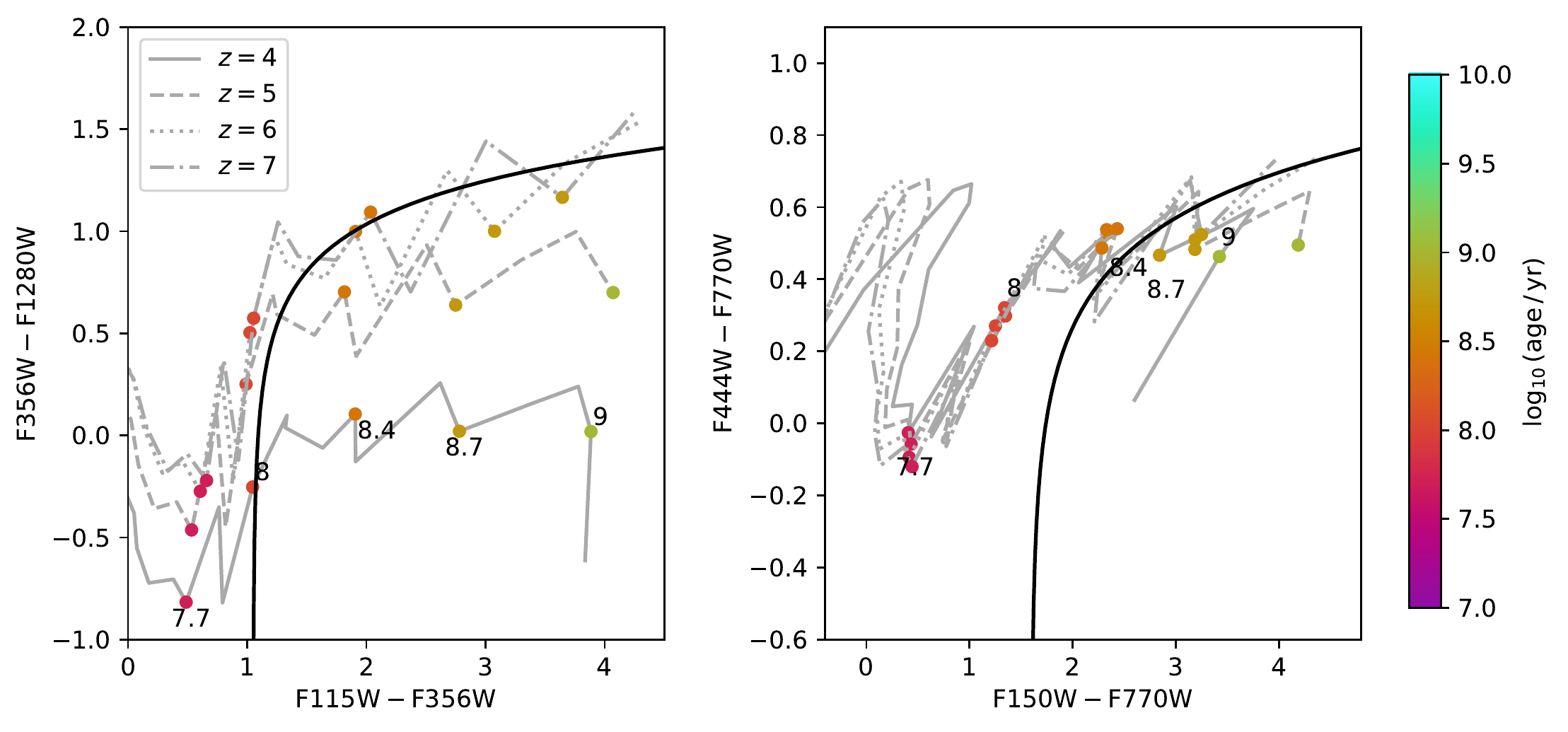}
    \caption{
        Theoretical tracks in colour -- colour space for an instantaneous burst (BPASS). 
        Each line shows a different observed redshift between $4 \leqslant z \leqslant 8$.
        Coloured points show where each population reaches a given age.
        Tracks are truncated where they reach the age of the Universe at that redshift.
        \textit{Left:} the $\mathrm{[F115W - F356W]}$ \& $\mathrm{[F356W - F1280W]}$ space.
        \textit{Right:} the $\mathrm{[F150W - F770W]}$ \& $\mathrm{[F444W - F770W]}$ space.
    }
    \label{fig:theoretical_colour_curves}
\end{figure*}

\subsection{Detectability with JWST}
\label{sec:jwst}

\fig{jwst_uvj_evolution} shows the redshift evolution of the rest--frame wavelength of various NIRCam and MIRI filters alongside the U, V and J filters.
At $z > 3$ only the shorter wavelength U and V filters are probed directly by NIRCam.
Where MIRI photometry is unavailable, in order to obtain the VJ colour templates are used to infer the J band photometry.
These are either empirical (typically derived at lower redshift) or taken from theoretical SPS models, which may suffer from systematic uncertainties and biases.
Even where MIRI photometry is available, interpolation of empirical data or templates can lead to uncertainties in derived rest-frame colours.
It is therefore worthwhile to explore the selection of passive galaxies using filters solely defined in the observer frame.
We focus on the NIRCam and MIRI instruments on JWST, which is already discovering passive galaxy candidates at very high redshifts \citep{carnall_surprising_2023,marchesini_early_2023}.

\fig{F115W_F356W_F356W_F1280W_ssfr_curve} shows two example two--colour spaces, using the F115W, F150W, F356W and F444W NIRCam filters and the F770W \& F1280W filters from MIRI.
Passive galaxies in \flares\ at $5 \leqslant z \leqslant 8$ are clearly separable in both of these spaces.
In the first, [F115W - F356W] / [F356W - F1280W], there is the clearest distinction between the passive and star forming populations.
Unfortunately, the sensitivity in F1280W is low (see \App{sensitivity}), leading to lower numbers of detected passive galaxies for which the colours can be measured.
In the second, using only the photometry that will be available in the upcoming COSMOS Web survey \citep{kartaltepe_cosmos-webb_2021}, [F150W - F770W] / [F444W - F770W], there is greater overlap, leading to lower purity, but the sensitivity in all of the bands is high, so more passive objects are detected.

\begin{table}
    \centering
    \caption{Parameters for the passive selection curves (equation \ref{eq:pass}) for two colour spaces, given in the text.}
    \label{tab:pass_params}
    \begin{tabular}{llcccc}
    \hline
        x & y & $y_0$ & $m$ & $x_0$ & $n$ \\
      \hline
        \thead{$\mathrm{[F150W -}$ \\ $\mathrm{F770W]}$} & \thead{$\mathrm{[F444W -}$ \\ $\mathrm{F770W]}$} & 6.49 & -6 & 1.6 & -0.04 \\
        \thead{$\mathrm{[F115W -}$ \\ $\mathrm{F356W]}$} & \thead{$\mathrm{[F356W -}$ \\ $\mathrm{F1280W]}$} & 7.1 & -6.0 & 1.49 & -0.03 \\
      \hline
    \end{tabular}
\end{table}

We draw polynomial selection curves in both colour spaces, with the following form
\begin{align} \label{eq:pass}
    y = y_0 + m (x - x_0)^n
\end{align}
and fit in order to optimise the F-score,
\begin{align}
    F = \rm \frac{2 \times TP}{2 \times TP + FP + FN}
\end{align}
where TP, FP and FN are the number of true positives, false positives and false negatives, respectively.
The fitted parameters\footnote{fit using \texttt{scipy.optimize.minimize}.} for both colour choices are provided in \tab{pass_params}.
This selection achieves relatively high completeness and purity of the passive galaxy population in both cases.
However, we caution that these colours are only calculated for \flares\ down to $z = 5$, and that lower redshift interlopers may exist which pollute the selection region.
It is also worth highlighting that the colour evolution of galaxies is highly sensitive to the age, due to the redshifting of emission lines in and out of filter bands, as demonstrated in \cite{wilkins_first_2022}.
We have carefully chosen the filters above to avoid this situation.
Finally, in \fig{theoretical_colour_curves} we show theoretical curves as a function of age for simple stellar populations at $z = [4, 5, 6, 7]$ using a single instantaneous burst with BPASS ($Z = 0.01$).
This further demonstrates the efficacy of our selection region in separating populations $\sim > 20$ Myr in age from their younger counterparts.
We explored a number of other colour combinations and found that the inclusion of longer wavelength photometry from MIRI (covering the rest--frame J band) achieves the best separation of passive galaxies from the star--forming galaxy population.

\section{Conclusions}
\label{sec:conclusions}

We have studied the high redshift passive galaxy populations in the \flares\ and \eagle\ simulations, and their detectability in rest-- and observer--frame photometry.
Our conclusions are as follows:

\begin{itemize}
    \item We computed the volume normalised number densities of massive ($M_{\star} \,/\, \mathrm{M_{\odot}} > 5 \times 10^{9}$) passive galaxies.
    At $z \leqslant 5$ we find that \eagle\ is in good agreement with observational constraints.
    Using the unique \flares\ simulation approach, we can extend the redshift range to the $z > 5$ regime, and we find passive populations up to $z \sim 8$ for certain intrinsic selection functions, with number densities of $\sim 10^{-5.5} \, \mathrm{Mpc^{-3}}$ at $z \sim 5$.
    \item We also compute surface number densities for various NIRCam flux cuts, and predict 85 passive objects per square degree at $z \geqslant 5$, for F200W > 27.5. For the upcoming COSMOS Web survey on JWST, we predict that $\sim 50$ passive objects at $z \geqslant 5$ will be detected.
    \item By looking at individual star formation and black hole accretion histories, as well as overall galaxy population demographics, we conclude that feedback from accretion onto supermassive black holes is the primary cause of passivity in these high redshift galaxies.
    The main motivation is the strong anti-correlation between star formation and black hole accretion history, the higher normalisation of the stellar mass--black hole relation, and the reduced gas fraction and star forming gas fraction in the passive population.
    \item Passive galaxies in \flares\ form earlier than those that are star forming, by $\sim 150$ Myr.
    Formation times in \eagle\ and \flares\ are in good agreement with recent observational estimates from HST \& JWST passive candidates at $z \leqslant 5$, however we do not predict the very early formation times ($z \sim 10$) of some observational candidates at these redshifts, instead finding passive galaxies that formed at these redshifts at $z = 6$.
    \item After forward modelling the emission from our galaxies, we find that none of our passive objects at $z = 5$ in \flares\ lie within the typical UVJ selection space, and very few at $z = 3$ in \eagle.
    This is due to the fact that, despite being defined as passive, these galaxies still have relatively young stellar population ages at these redshifts.
    This means that there is not enough time for UV colours to become red enough to reach the typical selection space.
    We present new selection regions in UVJ space for capturing passive galaxies at $z \geqslant 3$.
    \item We make predictions for observer--frame colour distributions from NIRCam and MIRI, and draw selection regions for passive galaxies at $z \geqslant 5$ that maximise the completeness and purity.
\end{itemize}

This is a very exciting time for the understanding and analysis of passive galaxies in the early Universe, due to a confluence of new observational data, as well as new models probing new regimes in volume and resolution.
Due to the rarity of these galaxies, the \flares\ simulation approach is absolutely necessary to capture the effective volume, as well as the highly overdense regions, where such massive passive objects are prevalent, producing sufficient numbers to enable a statistical analysis.
We have shown that passive galaxies at high redshift are a natural consequence of galaxy evolution models incorporating feedback from supermassive black holes.
However, it still remains challenging to simultaneously match the number densities of these passive objects as well as the abundance of dusty star forming galaxies, as traced by their sub-mm emission \citep{wang_multi-wavelength_2019,cowley_evolution_2019,lovell_reproducing_2021,vijayan_first_2021}.

If our predictions are correct, tens of passive galaxies at high redshift ($z \geqslant 5$) will soon be detected by JWST.
Future surveys on the Euclid and Roman observatories will cover much wider fields, potentially increasing the numbers of these passive objects significantly \citep{euclid_collaboration_euclid_2023}.
Detailed follow up with JWST and ALMA will help to uncover the causes of this passivity, confirming the picture painted here, or revealing other processes responsible for passivity in the early Universe.

\section*{Acknowledgements}
The authors wish to thank the anonymous referee for their insightful comments and suggestions that improved this manuscript. 
We also wish to thank Adam Carnall for providing their data, and Maximilien Franco, Caitlin Casey, James Trussler and James Trayford for helpful discussions.
We thank the \eagle\ team for their efforts in developing the \eagle\ simulation code.
We also wish to acknowledge the following open source software packages used in the analysis: \textsf{scipy} \citep{2020SciPy-NMeth}, \textsf{Astropy} \citep{robitaille_astropy:_2013} and \textsf{matplotlib} \citep{Hunter:2007}.

This work used the DiRAC@Durham facility managed by the Institute for Computational Cosmology on behalf of the STFC DiRAC HPC Facility (www.dirac.ac.uk).
The equipment was funded by BEIS capital funding via STFC capital grants ST/K00042X/1, ST/P002293/1, ST/R002371/1 and ST/S002502/1, Durham University and STFC operations grant ST/R000832/1.
DiRAC is part of the National e-Infrastructure. The \eagle\ simulations were performed using the DiRAC-2 facility at Durham, managed by the ICC, and the PRACE facility Curie based in France at TGCC, CEA, Bruyeres-le-Chatel.

CCL acknowledges support from a Dennis Sciama fellowship funded by the University of Portsmouth for the Institute of Cosmology and Gravitation, and the Royal Society under grant RGF/EA/181016.
PAT 
acknowledges support from the Science and Technology Facilities Council (grant number ST/P000525/1).
DI acknowledges support by the European Research Council via ERC Consolidator Grant KETJU (no. 818930).
The Cosmic Dawn Center (DAWN) is funded by the Danish National Research Foundation under grant No. 140.

We list here the roles and contributions of the authors according to the Contributor Roles Taxonomy (CRediT)\footnote{\url{https://credit.niso.org/}}.
\textbf{Christopher C. Lovell}: Conceptualization, Data curation, Methodolgy, Investigation, Formal Analysis, Writing - original draft.
\textbf{Will Roper, Aswin P. Vijayan, Louise Seeyave, Dimitrios Irodotou}: Formal Analysis, Visualization, Writing - review \& editing.
\textbf{Stephen M. Wilkins}: Conceptualization, Formal Analysis, Visualization.
\textbf{Peter Thomas}: Resources, Writing - review \& editing.
\textbf{Jussi Kuusisto, Flaminia Fortuni, Emiliano Merlin, Paola Santini}: Writing - review \& editing.

\section*{Data Availability}
The data underlying this article (stellar masses and star formation rates between $z = 5-10$) are available at \href{https://flaresimulations.github.io/data.html}{flaresimulations.github.io/data}.
Additional data is available upon reasonable request to the corresponding author.
All of the codes used for the data analysis are public and available at \href{https://github.com/flaresimulations}{github.com/flaresimulations}.

\bibliographystyle{mnras}
\bibliography{flares_passive,custom}



\appendix
\section{Effect of Aperture on Derived Colours}
\label{sec:aperture_UVJ}

\begin{figure}
    \centering
    \includegraphics[width=\columnwidth]{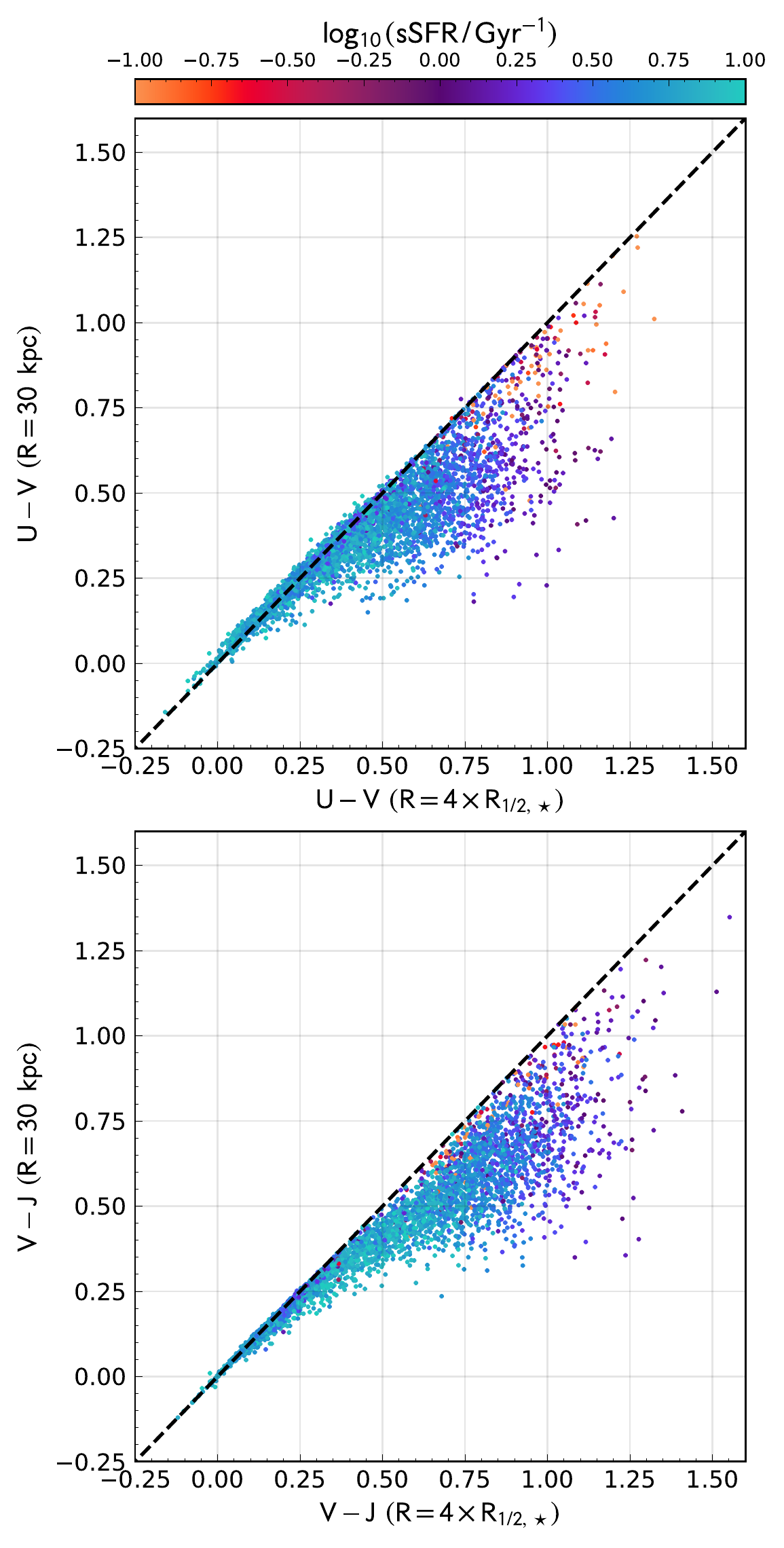}
    \caption{UV and VJ colours obtained using a 3D aperture with radius 30 kpc, and one defined using the half-mass radius, $R = 4 \times R_{1/2,\star}$.
    }
    \label{fig:UVJ_aperture}
\end{figure}

As discussed in \sec{forward_modelling}, we adopt the same forward modelling pipeline as presented in \cite{wilkins_nebular-line_2020,vijayan_first_2021}, with the only major modification being the 3D aperture over which the observable property is calculated.
In those previous works we adopted a fixed aperture with radius 30 pkpc.
In this work, to match the aperture used to define intrinsic properties (see \sec{aperture_timescale}), we calculate observed properties within a 3D aperture with radius $R = [1, 3, 5, 10, 20, 30, 40, 50, 70, 100] \mathrm{pkpc}$ closest to $R = 4 \times R_{1/2,\star}$.
\fig{UVJ_aperture} shows the affect of this assumption on the UV and VJ colours.
For the majority of galaxies, using the modified aperture leads to redder colours, by up to 0.8 magnitudes, and the size of this effect is dependent on the star formation rate within the aperture.
This highlights the importance of carefully comparing like-for-like properties in both the observations and simulations.

\section{Sensitivity and Detection Fractions}
\label{sec:sensitivity}

\begin{figure}
    \centering
    \includegraphics[width=\columnwidth]{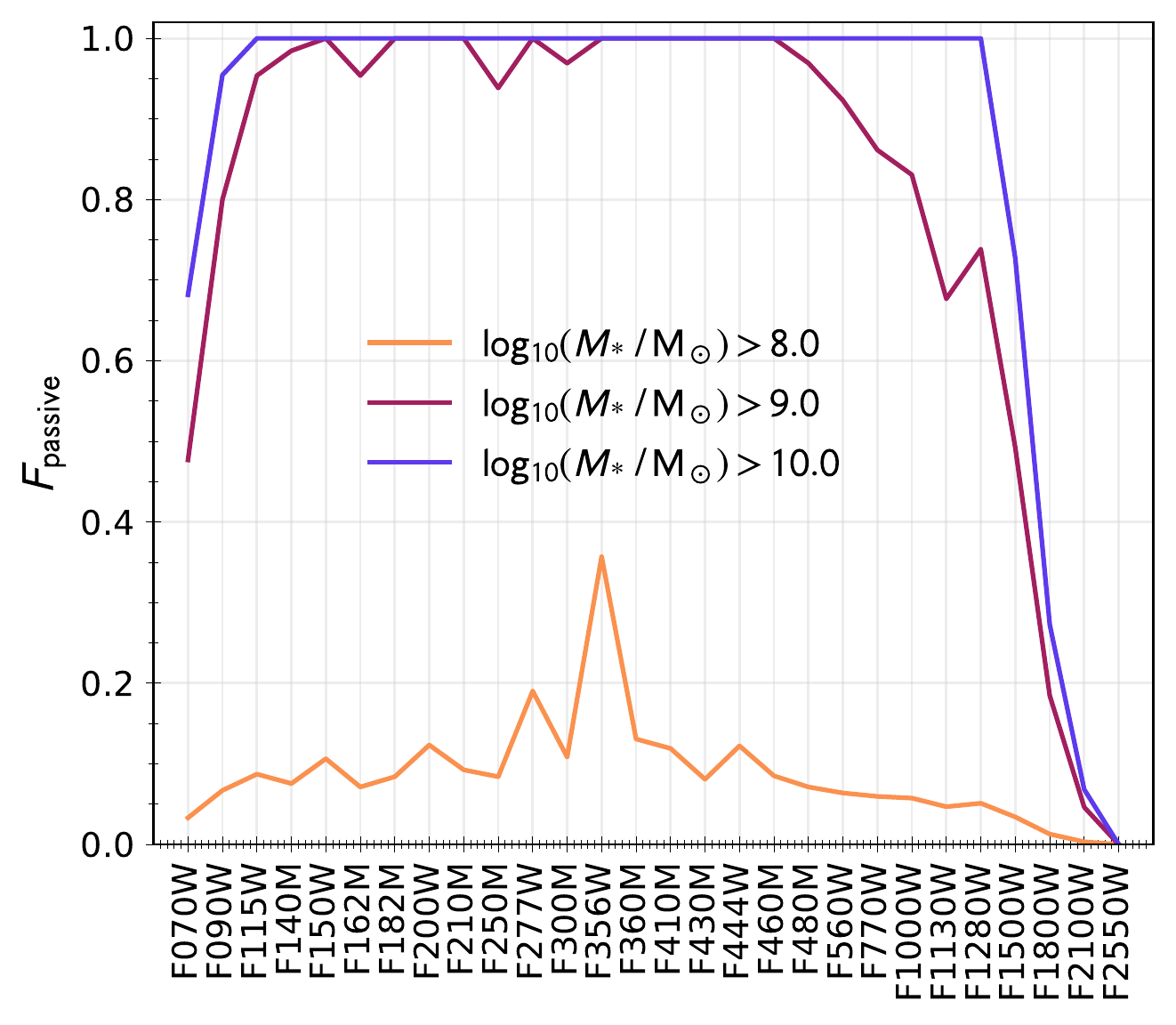}
    \caption{Fraction of all passive galaxies ($\mathrm{sSFR < -1 / Gyr^{-1}}$) detected in each JWST NIRCam and MIRI band.
    }
    \label{fig:passive_fraction_filter}
\end{figure}

The sensitivity of the NIRCam and MIRI instruments on JWST will limit the detectability of passive galaxies in certain bands.
We use the sensitivities provided online\footnote{NIRCam: \href{https://jwst-docs.stsci.edu/jwst-near-infrared-camera/nircam-performance/nircam-sensitivity}{https://jwst-docs.stsci.edu/jwst-near-infrared-camera/nircam-performance/nircam-sensitivity}, MIRI: \href{https://jwst-docs.stsci.edu/jwst-mid-infrared-instrument/miri-performance/miri-sensitivity}{https://jwst-docs.stsci.edu/jwst-mid-infrared-instrument/miri-performance/miri-sensitivity}} to assess this given the predicted emission in our \textsc{Flares} galaxies, whilst noting that these sensitivity limits are subject to change post instrument commissioning.
\fig{passive_fraction_filter} shows the fraction of passive galaxies detected for different stellar mass limits in each band.
The most massive galaxies are all detected in almost all bands, except the longest wavelength bands in the mid-infrared, and the shortest wavelengths probing the optical.

\bsp	
\label{lastpage}
\end{document}